%% file: main.tex
\documentclass[conference]{IEEEtran}
\usepackage{cite}
\usepackage{amsmath,amssymb,amsfonts}
\usepackage{graphicx}
\usepackage{textcomp}
 
\usepackage[colorlinks,bookmarksopen,bookmarksnumbered,allcolors=black]{hyperref}
\usepackage{cite}
\usepackage{booktabs}
\usepackage{makecell}
\usepackage{multirow}
\usepackage{comment}
\usepackage{moresize}

\usepackage[dvipsnames]{xcolor}

\newcommand{\red}[1]{\textcolor{black}{#1}}

\definecolor{Tomato}{RGB}{255,99,71}  
\definecolor{NiceGreen}{RGB}{34,139,34}

\def\BibTeX{{\rm B\kern-.05em{\sc i\kern-.025em b}\kern-.08em
    T\kern-.1667em\lower.7ex\hbox{E}\kern-.125emX}}
\begin{document}

\title{
Multi-Objective Loss Balancing in Physics-Informed Neural Networks for Fluid Flow Applications 
}

\author{\IEEEauthorblockN{Afrah Farea}
\IEEEauthorblockA{\textit{CSE Department, Informatics Institute} \\
\textit{Istanbul Technical University}\\
Istanbul, Turkiye \\
farea16@itu.edu.tr}
\and
\IEEEauthorblockN{ Saiful Khan}
\IEEEauthorblockA{\textit{Scientific Computing} \\
\textit{Rutherford Appleton Laboratory, STFC}\\
Didcot, OX11 0QX, UK \\
saiful.khan@stfc.ac.uk }
\and
\IEEEauthorblockN{ Mustafa Serdar Celebi}
\IEEEauthorblockA{\textit{CSE Department, Informatics Institute} \\
\textit{Istanbul Technical University}\\
Istanbul, Turkiye \\
mscelebi@itu.edu.tr}
}

\maketitle

\input{sections/0_abstract}

\begin{IEEEkeywords}
Physics-Informed Neural Networks, PINN, Partial Differential Equations, PDE, Navier-Stokes, Multi-objective Loss, Loss Function.
\end{IEEEkeywords}

\input{sections/1_introduction}

\input{sections/2_3_background_related_work}

\input{sections/4_methodology}

\input{sections/5_results}
\input{sections/6_conclusion}

\section*{Acknowledgment}
We thank the National Center for High-Performance Computing of Turkey (UHeM) for providing computing resources under grant number 5010662021. 
We thank Dr. Emre Cenk Ersan and Dr. Reza Daryani from ITU for the helpful discussions and for generating the test cases using OpenFOAM software.  

\bibliographystyle{abbrv-doi-narrow}
\bibliography{bibliography}

\end{document}

%% file: sections/0_abstract.tex
\begin{abstract}
Physics-Informed Neural Networks (PINNs) have emerged as a promising machine learning approach for solving partial differential equations (PDEs). 
However, PINNs face significant challenges in balancing multi-objective losses, as multiple competing loss terms such as physics residuals, boundary conditions, and initial conditions must be appropriately weighted. 
While various loss balancing schemes have been proposed, they have been implemented within neural network architectures with fixed activation functions, and their effectiveness has been assessed using simpler PDEs. We hypothesize that the effectiveness of loss balancing schemes depends not only on the balancing strategy itself, but also on the loss function design and the neural network's inherent function approximation capabilities, which are influenced by the choice of activation function. In this paper, we extend existing solutions by incorporating trainable activation functions within the neural network architecture and evaluate the proposed approach on complex fluid flow applications modeled by the Navier-Stokes equations. 
Our evaluation across diverse Navier-Stokes problems demonstrates that this proposed solution achieves root mean square error (RMSE) improvements ranging from 7.4\% to 95.2\% across different scenarios. 
These findings highlight the importance of carefully designing the loss function and selecting activation functions for effective loss balancing.
\end{abstract}

%% file: sections/1_introduction.tex
\section{Introduction}
\label{sec:introduction}

The Navier-Stokes equations are a system of partial differential equations (PDEs) that govern the motion of viscous fluids, capturing the complex interplay between momentum conservation, mass conservation, and viscous effects. 
Despite being one of the most challenging systems to solve in fluid mechanics, these equations are essential for understanding fluid behavior across diverse applications ranging from aerodynamics and biomedical flows to climate modeling and industrial processes in both engineering and scientific contexts~\cite{temam2024navier, pope2001turbulent}.
Numerical computational fluid dynamics (CFD) approaches to solve Navier-Stokes equations, including finite element and spectral methods, often require high-performance computing (HPC) infrastructure. These methods require extensive domain discretizations, sophisticated mesh generation for complex geometries, and careful treatment of boundary conditions, making them computationally expensive and sometimes prohibitive for inverse problems or large real-time predictions \cite{ballarin2017numerical}. Additionally, traditional CFD methods struggle with sparse or noisy data~\cite{brunton2017discovering, wang2018propagation, reinbold2021robust}, unknown boundary conditions~\cite{feigley2011deriving, xu2018assessment}, or scenarios requiring simultaneous forward simulation and parameter identification~\cite{sun2020surrogate}.
 
Physics-informed neural networks (PINNs) have emerged as a promising alternative for solving PDEs, including the Navier-Stokes equations. PINNs offer a mesh-free framework that can simultaneously tackle forward and inverse problems within a unified optimization approach. PINNs have demonstrated effectiveness, particularly in fluid flow applications, from simple channel flows to complex simulations and turbulent flows~\cite{jin2021nsfnets, mao2020physics, farea2025:fsi-ibm, farea2025:qcpinn}.
Despite their promise, PINNs remain at the research stage due to significant limitations, including training instability, difficulty handling complex geometries, poor scalability to high-dimensional problems, and challenges with multi-scale physics, among others. Ongoing research efforts are addressing these fundamental issues.
One area of research focuses on modeling the loss function, as the PINN loss function typically comprises multiple competing terms, including initial conditions, boundary conditions, and physics loss terms.
For instance, fluid flow problems involve the Navier-Stokes equations (momentum and continuity equations), various boundary conditions (such as no-slip, inlet, outlet, and wall conditions), and initial conditions. Each of these objectives operates at different physical scales and exhibits varying degrees of optimization difficulty, creating a complex multi-objective landscape where improper loss balancing can lead to poor convergence or physically unrealistic solutions.

Numerous multi-objective loss balancing schemes have been proposed for PINNs, with the most prominent approaches including Residual-Based Attention (RBA)~\cite{anagnostopoulos4586276residual}, Self-Adaptive (SA)~\cite{mcclenny2020self}, Learning Rate Annealing (LRA)~\cite{wang2021understanding}, and Gradient Normalization (GradNorm)~\cite{chen2018gradnorm} for solving PDEs. However, these schemes have primarily been evaluated on simpler benchmark problems rather than complex fluid flow scenarios. There is considerable potential to expand this research to include more complex PDEs, such as the Navier-Stokes equations. This could involve examining canonical fluid flow problems, including lid-driven cavity flow, plane Poiseuille flow between parallel plates, and blood flow simulations that account for both slip and no-slip boundary conditions.
Furthermore, existing multi-objective loss balancing implementations have predominantly relied on fixed activation functions, such as the Tanh, limiting our understanding of how a trainable activation function might influence balancing effectiveness. 
We propose that the effectiveness of multi-objective loss balancing for PINNs is influenced not only by the balancing methods themselves but also by the neural network architecture's inherent function approximation capabilities. These capabilities are affected by the choice of activation function and the design of the loss function. 
We address these limitations through two key contributions: 

\noindent (1) We extend established multi-objective loss balancing schemes by incorporating the SiLU activation function and trainable B-spline basis functions~\cite{liu2024kan}. We then compare their performance against existing implementations that use a fixed activation function, specifically Tanh.
    
\noindent (2)
We provide a comprehensive evaluation of these loss balancing strategies applied to various Navier-Stokes fluid flow problems, analyzing how the interplay between activation functions and loss balancing impacts solution quality across different fluid dynamics scenarios.

The source code and pre-trained models used in this study are publicly available on GitHub at~\url{https://github.com/afrah/pinn_adaptive_weighting} to facilitate reproducibility and further exploration.

%% file: sections/2_3_background_related_work.tex
\section{Background}
\label{sec:problem}

Incompressible Navier-Stokes equations can be stated in a general form:

{\small
\begin{align}
    \frac{\partial \mathbf{u}}{\partial t} + (\mathbf{u} \cdot \nabla)\mathbf{u} & + \frac{1}{\rho}\nabla p - \nu \nabla^2 \mathbf{u} = \mathbf{f}(\mathbf{x}, t),  \quad t \in [0, T], \quad \mathbf{x} \in \Omega \label{eq_generalPDE} \\
    \nabla \cdot \mathbf{u} &= 0, \quad t \in [0, T], \quad \mathbf{x} \in \Omega \notag \\
    \mathcal{B}_k[\mathbf{u}, p] &= \mathbf{g}_k(\mathbf{x}, t), \quad t \in [0, T], \quad \mathbf{x} \in \Gamma_k \subset \partial \Omega \notag
\end{align}
}
\noindent where $\mathcal{B}_k$ denotes boundary operators for velocity and pressure conditions, $\mathbf{f}(\mathbf{x}, t)$ is the source term or the external force, $\mathbf{g}_k(\mathbf{x}, t)$ are the boundary conditions, $\nu$ is the viscosity, and $\rho$ is the fluid density.

The physics loss ensures that the neural network solution captures knowledge about the governing PDE equation by minimizing the residual of the differential equation on a set of randomly selected representative points. The boundary and initial conditions ensure the uniqueness of the solution to the PDE problem. Together, these terms define a multi-objective optimization problem (MOOP)~\cite{lobato2017multi,maddu2022inverse}, which is commonly addressed by combining them into a single aggregated loss function:

{\small
\begin{align}
    \mathcal{L}(\theta) &= arg\underset{\theta}{min}\sum \limits_{k=0}^n  \mathcal{L}_k(\theta) \label{eq_pinn_no_weight}\\
    &= arg\underset{\theta}{min}\Big( \mathcal{L}(\mathcal{D}[u(t,x);\alpha]  - f(t,x) ) \notag  \notag \\
    &+  \sum \limits_{k=1}^{n_b}  \mathcal{L}(\mathcal{B}_k[u(t,x)] - g_k(t,x))  \Big) \notag  \notag 
\end{align} 
 }
\noindent where $n$ is the number of terms in the loss function. 

Eq.~\ref{eq_pinn_no_weight} shows the multi-objective constraints of the PDE Eq.~\ref{eq_generalPDE} reduced to a single constraint. This only applies under the assumption that a very low loss can be reached on each term.
Therefore, this soft approach to imposing boundary condition constraints is not guaranteed, as the objective is to minimize the loss function as a whole. 

The simplest approach to balancing the different parts of the loss function is multiplying each term by a weighting coefficient $\lambda_k$ during the training process as follows: 

{\small
\begin{align}
    \mathcal{L}(\theta) &= arg\underset{\theta}{min}\sum \limits_{k=0}^n  \lambda_k \mathcal{L}_k(\theta)  \label{eq:pinn_wloss} \\
    &=arg\underset{\theta}{min}\Big(\lambda_0 \mathcal{L}(\mathcal{D}[u(t,x);\alpha]  - f(t,x) )  \notag \\
    &+  \sum \limits_{k=1}^{n_b} \lambda_k \mathcal{L}(\mathcal{B}_k[u(t,x)] - g_k(t,x))  \Big) \notag 
\end{align}
}
\noindent where $\lambda_k$ is the weighting coefficient of the loss term. The loss function $\mathcal{L}$ is typically chosen as the mean squared error (MSE).

Many challenges may arise with the PINN loss function defined in Eq.~\ref{eq:pinn_wloss}, particularly when dealing with nonconvex problems. These challenges include, but are not limited to:

\vspace{2mm}\noindent
(a) Challenge of Physics-Based Regularization:
A key observation in PINN loss function is that it often includes terms with different physical scales. This is because the physics loss may involve high-order derivatives, which can be ill-conditioned and negatively affect the quality of training~\cite{basir2022critical,basir2023investigating}. Meer et al.~\cite{van2022optimally} showed that the loss function for solving any linear PDE is convex and thus does not have local minima; however, for nonlinear PDEs, the loss landscape is generally nonconvex with no convergence guarantees.

\vspace{2mm}\noindent
(b) Local Minima and Saddle Points: 
Nonconvex problems have multiple local minima and potentially many saddle points. 
This makes optimization far more challenging, as gradient-based methods used for training can converge to suboptimal solutions or become trapped in ill-conditioned areas of the landscape~\cite{dauphin2014identifying}. Furthermore, the mean Squared Error (MSE) used for the loss function does not provide any inherent mechanism to escape these possibly suboptimal points. As a result, the model may converge to a poor local minimum~\cite{liu2021non}.

\vspace{2mm}\noindent
(c) Slow Convergence:
During training, gradients may become extremely small, leading to slow convergence. Moreover, standard multi-layer perceptrons (MLPs) exhibit spectral bias, meaning they tend to learn low-frequency components of the solution more effectively while struggling to capture high-frequency features~\cite{rahaman2019spectral,tancik2020fourier}. To address these limitations, a local B-spline parameterization combined with SiLU activation functions can provide greater flexibility in function approximation~\cite{farea2025:learnable, farea2025:fsi-ibm}.

\vspace{2mm}\noindent
(d) Difficulty in Balancing Loss Components:
Selecting weight values is essential to avoid failure and accelerate convergence.
However, a manual assignment can make the solution quality dependent on tedious hyperparameter tuning, especially with many loss terms. If some gradient parts are small while others are large, training will not be effective for the small gradient parts. 

\section{Related Work}

Various methods have been proposed to address the loss balancing issues in PINNs, using different criteria such as temporal dataset information~\cite{krishnapriyan2021characterizing,wight2020solving,wang2022respecting}, gradient information~\cite{wang2021understanding,vemuri2023gradient,wang2022and}, loss term history statistics~\cite{Heydari2020,bischof2021multi}, training dataset information~\cite{mcclenny2020self,li2022dynamic,sahli2020physics}, direct imposition of boundary conditions~\cite{ren2022phycrnet}, maximum likelihood estimations~\cite{xiang2022self}, Lagrangian relaxation methods~\cite{son2023enhanced} and others~\cite{jagtap2020adaptive,ji2021stiff,moseley2023finite}. 

In the following, we review the loss balancing schemes we selected in this study (based on their adoption, citation, and suitability):

\vspace{2mm}\noindent
\textbf{Residual-Based Attention (RBA):} 
This method~\cite{anagnostopoulos4586276residual} updates weights using the normalized residual magnitudes according to $\lambda^{(k+1)}_i = \gamma\lambda^k_i + \eta^* |e_i| / ||e||_\infty$, where $\gamma$ and $\eta^*$ are user-defined values representing the decay parameter and the learning rate, respectively. In this study, we choose 0.5 for both $\gamma$ and $\eta^*$ to avoid abrupt changes in the lambda values. RBA operates on batch-wise residuals and has minimal computational overhead since it only requires computing residual norms, which have already been calculated during training.

\vspace{2mm}\noindent
\textbf{Self-Adaptive (SA)}: This is a gradient-based approach~\cite{mcclenny2020self} which uses learnable weight parameters that are updated via gradient ascent with $\nabla_\lambda L = \frac{1}{2} m'(\lambda) |\text{residual}|^2$, where $m(\lambda)$ is a mask function and $m'$ is its derivative. In the present work, we choose a polynomial mask function with power 2: $m(\lambda) = \lambda^2$ with learning rate $\alpha_{SA} = 0.001$. SA requires gradient computation for each weight parameter, resulting in moderate computational overhead.
For both RBA and SA, the loss terms are computed point-wise rather than using mean squared error, allowing for individual sample weighting within each batch.

\vspace{2mm}\noindent
\textbf{Learning Rate Annealing (LRA):} This is a gradient-based method~\cite{wang2022and} which computes the ratio of maximum residual gradients to individual loss gradients: $\hat{\lambda}_i = \max_{\theta}|\nabla_\theta L_r(\theta)| / |\nabla_\theta L_i(\theta)|$, then updates weights using exponential moving average: $\lambda_i = (1-\alpha)\lambda_i + \alpha\hat{\lambda}_i$. Similar to RBA, we choose 0.5 for $\alpha$ to avoid abrupt changes in the lambda values. LRA has a moderate computational cost due to gradient norm computations for each loss term.

\vspace{2mm}\noindent
\textbf{Gradient Normalization (GradNorm):} This method~\cite{chen2018gradnorm} balances multiple objectives by equalizing gradient magnitudes across tasks. It computes target gradient norms $G_W(t) \times [r_i(t)]^\alpha$ where $r_i(t)$ represents relative inverse training rates, $\alpha$ controls the asymmetry parameter, and $G_W(t)$ is the average gradient norm across all tasks at training step $t$. GradNorm has the highest computational overhead as it requires computing gradients for each task, gradient norms, and solving an additional optimization problem for weight updates.

The above-mentioned loss balancing schemes were implemented using traditional MLP neural networks with fixed activation functions. Additionally, those studies focused on simpler PDEs. Our work builds on this foundation by incorporating the SiLU activation function and learnable B-spline basis functions. We also examined more complex PDEs, such as the Navier-Stokes equations, under various boundary and initial conditions.

%% file: sections/4_methodology.tex
\section{Methodology}

%------------------------------------------------------------------------------%
% NOTE: This figure is from methodology. Moved here to fix spacing.
\begin{figure*}[t]
    \centering
    \includegraphics[width=1.0\textwidth]{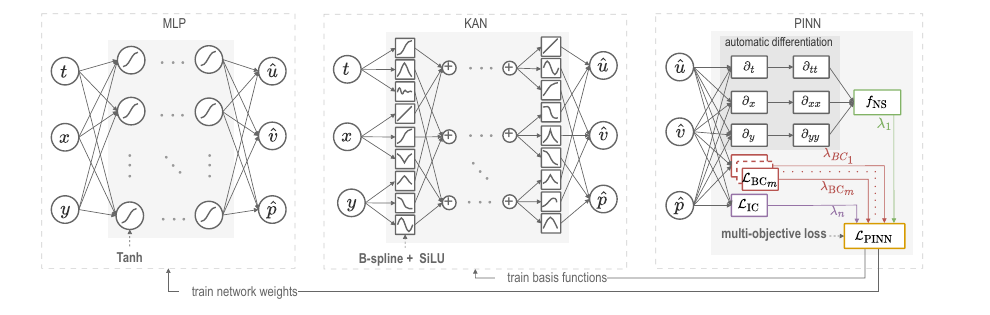}
    \caption{
    The diagram illustrates two neural network architectures: the Tanh activation function is applied in a traditional multi-layer perceptron (MLP), while SiLU+B-Spline is utilized in Kolmogorov-Arnold Networks (KAN). The predictions of these networks are fed to a PINN.
    The PINN's loss function includes Navier-Stokes loss ($f_{\text{NS}}$), $m$ boundary condition losses ($\mathcal{L}_{\text{BC}_1} \dots \mathcal{L}_{\text{BC}_m}$), and an initial condition loss ($\mathcal{L}_{\text{IC}}$). 
    As a result, the total loss becomes a multi-objective optimization problem, expressed as
    $ \mathcal{L}_{\text{PINN}} = 
    \textcolor[HTML]{82B366}{\lambda_1}f_{\text{NS}} + 
    \textcolor[HTML]{B85450}{\lambda_{\text{BC}_1}} \mathcal{L}_{\text{BC}_1} + \dots + 
    \textcolor[HTML]{B85450}{\lambda_{\text{BC}_m}} \mathcal{L}_{\text{BC}_m} + 
    \textcolor[HTML]{9673A6}{\lambda_n} \mathcal{L}_{\text{IC}}$. 
    The weighting factors,
    $\textcolor[HTML]{82B366}{\lambda_1}$, 
    $\textcolor[HTML]{B85450}{\lambda_{\text{BC}_1}} \dots \textcolor[HTML]{B85450}{\lambda_{\text{BC}_m}}$, and 
    $\textcolor[HTML]{9673A6}{\lambda_n}$ 
    are also treated as trainable parameters, and their optimal values are learned during training, along with the neural network weights.
    }
    \label{fig:architecture_diagram}
\end{figure*}
%------------------------------------------------------------------------------%

In this section, we discuss the neural network architectures designed to incorporate trainable activation functions and various use cases involving Navier-Stokes to evaluate the proposed solution. 
We also discuss how to design the PINN loss for each specific use case.

\subsection{Neural Network Architectures}

Fig.~\ref{fig:architecture_diagram} illustrates two neural network architectures. To evaluate the existing method, we implemented the fixed activation function, specifically the Tanh function, within a multilayer perceptron (MLP). In contrast, to assess our proposed approach, we incorporated a trainable activation function (a combination of B-spline and SiLU) into the Kolmogorov-Arnold Networks (KAN)~\cite{liu2024kan}.
The PINN model receives output from the preceding network and utilizes automatic differentiation from the PyTorch computational graph to compute derivatives and loss components. The loss components are multiplied by trainable weights using various loss balancing schemes.

The network structure in KANs learns a combination of simpler functions by decomposing a multivariate function $f(x)$ using the Kolmogorov-Arnold representation theorem~\cite{kolmogorov1957representation} such that:

{\small
\begin{equation}
    f(\mathbf{x}) = \sum_{q=0}^{2n} \Phi_q\left(\sum_{p=1}^{n} \phi_{q,p}(x_p)\right)
\end{equation}
}
\noindent where $\phi_{q,p}: [0,1] \rightarrow \mathbb{R}$  and $\Phi_q: \mathbb{R} \rightarrow \mathbb{R}$ are univariate functions. In practice, KANs implement a layer-wise approximation where each layer applies univariate transformations~\cite{liu2024kan}. Each KAN layer combines a base activation with learnable B-spline functions:

{\small
\begin{align*}
    \phi(x) &= \lambda_{\text{b}} \text{silu}(x) + \lambda_{\text{s}} \text{B-spline}(x) = \lambda_{\text{b}} \text{silu}(x) + \lambda_{\text{s}}  \sum_{i=1}^{g+d} c_i B_i^d(x)
\end{align*}
}
\noindent where $g$ is the grid size, $d$ is the spline order, and $\lambda_{\text{b}}$, $\lambda_{\text{s}}$ and $c_i$ are trainable coefficients. The B-spline basis functions $B_i^d(x)$ are defined recursively using the Cox-de Boor formula~\cite{de1972calculating}:

{\small
\begin{align*}
B_i^d(x) &= \frac{x - \xi_i}{\xi_{i+d} - \xi_i} B_i^{d-1}(x) + \frac{\xi_{i+d+1} - x}{\xi_{i+d+1} - \xi_{i+1}} B_{i+1}^{d-1}(x)
\end{align*}
}
\noindent with the base case for $d = 0$:

{\small
\begin{align*}
B_i^0(x) &= \begin{cases} 1, & \text{if } \xi_i \leq x < \xi_{i+1} \\0, & \text{otherwise} \end{cases}
\end{align*}
}

\red{
Our selection of SiLU and B-spline basis functions, as opposed to other options, is primarily influenced by the findings in the study by Farea et al.~\cite{farea2025:learnable}. SiLU is a smooth, globally defined activation function that effectively captures general nonlinear trends while maintaining stable gradient flow during backpropagation. On the other hand, B-spline basis functions offer localized support through piecewise polynomial construction, which allows for precise representation of sharp gradients, boundary layers, and discontinuities that frequently arise in fluid dynamics problems studied in this work.
}

\subsection{Use Cases: Fluid Flow PDEs}
\label{sec:use-cases}

We evaluate the performance of the developed PINN models in solving various laminar flow cases. The results are then compared with numerical solutions obtained using the finite volume method, implemented through OpenFOAM’s C++ finite volume library.

All the cases examined involve incompressible flow, modeled using the time-dependent two-dimensional Navier-Stokes equations. These problems display varying levels of stiffness and multi-scale behavior. The diversity of boundary conditions, including Dirichlet, Neumann, and mixed types, along with the different physical scales across the cases, where the input to the neural network represents the temporal and the spatial coordinates, and the outputs are the predicted velocity components and the pressure values, as shown in Fig.~\ref{fig:architecture_diagram}.

\vspace{2mm}\noindent
\textbf{(1) 2D Lid-driven Cavity Flow:}
The classical time-dependent steady-state flow in a 2D lid-driven cavity is a well-known benchmark problem, where one boundary moves with a constant velocity while the other boundaries are fixed. The governing equations are:

{\small
\begin{align}
        u_t + \mathbf{u . \nabla u} + \frac{1}{\rho}\nabla p -\nu \mathbf{\nabla^2 u} &=  0 \qquad \text{in} \quad \Omega    \label{eq:cavity1} \\
        \mathbf{ \nabla . u} &= 0  \quad \Omega_0 \notag \\
        u(t_0, x , y) =  v(t_0, x , y)=  p(t_0, x , y) &= 0    \quad \text{initial}\notag \\
        u(t , x , y_r) &= 1  \quad \Gamma_1 \quad  \text{up} \notag \\
        u(t , x_0 , y) =  u(t , x_r , y) = u(t ,x, y_0) &= 0 \quad \Gamma_0 \quad  \text{other boundaries} \notag
\end{align}
}
\noindent where the computation domain  $\Omega$ is a two-dimensional square cavity $\Omega= (0, 1) \times (0, 1)$.  Uniform discretization is used with grid $(N_x , N_y) = (100 , 100)$ and 10 seconds total simulation time with 0.01s time interval. $\Gamma_1$ is the top Dirichlet boundary condition of the cavity with velocity tangent to this side, $\Gamma_0$ denotes the other three stationary sides, and $\rho$ and $\nu$ are the fluid density and viscosity set to $1056 kg/m^3$ and $0.01kg/(m.s)$, respectively.

\vspace{2mm}\noindent
\textbf{(2) Plane Poiseuille Flow:}
The problem describes a laminar flow between two Infinite plates. 
The governing equations are:

{\small
\begin{align}
        \mathbf{u}_t + \mathbf{u} \cdot \nabla \mathbf{u} + \frac{1}{\rho}\nabla p - \nu \nabla^2 \mathbf{u} &= 0 \qquad \text{in} \quad \Omega \label{eq:poiseuille1_loss} \\
        \nabla \cdot \mathbf{u} &= 0 \qquad \text{in} \quad \Omega \notag \\
        u(t_0, x, y) &= {0.2}  \quad \text{initial} \notag \\
        v(t_0, x, y) =  p(t_0, x, y) &= 0 \quad \text{initial}  \notag \\
        u(t, x_0, y) &= {0.2} \quad \text{inlet} \notag \\
        \frac{\partial p}{\partial x}(t,x_0, y) &= 0  \quad \text{inlet} \notag \\
        p(t, x_{\text{r}}, y) &= 0  \quad \text{outlet} \notag \\
        \frac{\partial u}{\partial x}(t,x_r,y) =  v_x(t,x_r,y) &= 0 \quad \text{outlet}\notag \\
        \frac{\partial p}{\partial y}(t,x, y_{\pm r}) &= 0  \quad \text{p zero grad on the wall}\notag 
\end{align}
}
Uniform discretization is used for this case with regular grid $(N_x, N_y) = (62, 2001)$ where $(t,x,y) \in (t_0,t_r) \times (x_0,x_r)\times (y_0,y_r) =[0, 5]\times[0,1]\times[-0.0075, 0.0075]$ with $100$ time steps for the 5-second simulation. 
The kinematic viscosity is $3.3144*10^{-6}m^2/s$, and the density is $1060 kg/m^3$. 

This case presents several challenges for PINNs for several reasons: (1) The velocity component ($v$) is approximately two orders of magnitude smaller ($v\sim 10^{-2}$ m/s) than the streamwise velocity ($u\sim 0.2$ m/s), leading to multi-objective optimization imbalances where the network struggles to accurately capture both velocity scales simultaneously; (2) the computational domain spans only 15 mm in the y-direction ($y \in [-0.0075, 0.0075]$m) while extending 1m in the x-direction, creating an aspect ratio of approximately 67:1 that challenges spatial discretization and gradient computation; (3)  the parabolic velocity profile requires accurate representation of sharp velocity gradients near walls ($|y| = 0.0075$m) while maintaining smooth transitions in the channel center. This dual requirement makes it particularly challenging for the models to capture both the bulk flow and near-wall behavior simultaneously.

\vspace{2mm}\noindent
\textbf{(3) 2D Blood-flow in a tube with slip walls (BFS-Slip):}
This problem is a Newtonian, incompressible blood flow simulation in a tube with the governing equations as follows:

{\small
\begin{align}
        \mathbf{u}_t + \mathbf{u} \cdot \nabla \mathbf{u} + \frac{1}{\rho}\nabla p - \nu \nabla^2 \mathbf{u} &= 0 \qquad \text{in} \quad \Omega \label{eq:bfs_slip1}\\
        \nabla \cdot \mathbf{u} &= 0 \qquad \text{in} \quad \Omega \notag \\
        u(t_0, x, y)   &= {0.2}   \quad \text{initial} \notag \\
        v(t_0, x, y)  &= 0    \quad \text{initial} \notag \\
        p(t_0, x, y)  &= 0    \quad \text{initial} \notag \\
        u(t, x_0, y)  &= {0.2}   \quad \text{inlet} \notag \\
        \frac{\partial p}{\partial x}(t,x_0,y)  &=  0   \quad \text{p zero grad at inlet} \notag \\
        p(t, x_r, y) &= 0    \quad \text{outlet}\notag \\
        \frac{\partial u}{\partial x}(t,x_r,y) = v_x(t,x_r,y) &= 0    \quad \mathbf{u}\text{ zero grad at outlet} \notag \\
        \frac{\partial p}{\partial y}(t,x,y_{\pm r}) &=  0   \quad \text{p zero grad on the wall} \notag
\end{align}
}
\noindent We describe the PINN approach for approximating the solution with $(t,x,y) \in (t_0, t_r)\times(x_0,x_r)\times (y_0, y_r) =[0,5]\times [0, 1] \times [-7.5*10^{-3}, 7.5*10^{-3}]$. Viscosity = $0.00345 m^2/s$ and density $=$ $1056kg/m^3$ with initial and inlet velocity 0.2m/s in the x-direction and 0.0m/s in the y-direction. 
The inner fluid domain includes 340000 hexa-cells, while the wall surface has 40000 quad-cells. Cell-center locations are aligned in the x-direction.

\vspace{2mm}\noindent
\textbf{(4) 2D Blood-flow in a Tube with No-Slip Walls (BFS-No-Slip):}
The governing equations are:

{\small
\begin{align}
        \mathbf{u}_t + \mathbf{u} \cdot \nabla \mathbf{u} + \frac{1}{\rho}\nabla p - \nu \nabla^2 \mathbf{u} &= 0 \qquad \text{in} \quad \Omega \label{eq:no_slip_loss} \\
        \nabla \cdot \mathbf{u} &= 0 \qquad \text{in} \quad \Omega \notag \\
        u(t_0, x, y) &= {0.2}  \quad \text{initial} \notag \\
        v(t_0, x, y) =  p(t_0, x, y) &=0 \quad \text{initial}  \notag \\
        u(t, x_0, y) &= {0.2} \quad \text{inlet} \notag \\
        v(t, x_0, y) &= {0} \quad \text{inlet} \notag \\
        \frac{\partial p}{\partial x}(t,x_0, y) &= 0  \quad \text{inlet} \notag \\
        p(t, x_{\text{r}}, y) &= 0  \quad \text{outlet} \notag \\
        \frac{\partial u}{\partial x}(t,x_r,y) =  \frac{\partial v}{\partial x}(t,x_r,y) &= 0 \quad \text{outlet}\notag \\
        u(t,x, y_{\pm r}) = v(t,x, y_{\pm r}) &=0  \quad \text{no-slip on the wall}\notag \\
        \frac{\partial p}{\partial y}(t,x, y_{\pm r}) &= 0  \quad \text{p zero grad on the wall}\notag 
\end{align}
} 
\noindent Similar to the BFS-slip case, the inner fluid domain comprises 340,000 hexa-cells, while the wall surface has 40,000 quad-cells. Cell-center locations are aligned in the x-direction.

\subsection{Loss Function Design}

The loss function design varies for each fluid flow case based on the specific boundary conditions and physics involved. 
Not all boundary conditions are included in the loss function design due to several considerations.  While adaptive weighting schemes can dynamically adjust the relative importance of different loss terms based on training dynamics, they fundamentally operate under the assumption that all included loss terms contribute positively to solution quality. However, these methods do not address other issues, such as constraint redundancy or numerical incompatibility, between certain boundary conditions. 
Our approach involves \textbf{constraint selection}, where we choose which physics to enforce before \textbf{constraint weighting}, where loss balancing schemes adjust how strongly to enforce all physics. In other words, adaptive weighting optimizes the balance among included terms, while constraint selection optimizes which terms to include in the first place.

We prioritize boundary conditions that are essential for problem well-posedness while avoiding redundant or numerically problematic constraints. The selection criteria include: (a) boundary conditions required for the existence and unique solution according to PDE principles~\cite{evans2022partial}; 
(b) avoiding boundary conditions that create ill-conditioned optimization landscapes~\cite{wang2021understanding}. This occurs in some PDEs, such as the 2D wave equation, which may include Dirichlet and Neumann boundary conditions. Adding these two terms to the PINN loss function may result in a highly stiff or ill-conditioned optimization loss landscape, while only enforcing the Dirichlet boundary condition guarantees convergence with good approximation results. We justify that the reason is that the Naumann boundary condition contains a derivative that has different physical characteristics than the  “supervised” Dirichlet boundary conditions.
(c) prioritizing constraints that most directly govern the dominant physics~\cite{karniadakis2021physics}. For instance, in blood flow simulations, inlet conditions are more crucial than outlet conditions since the flow is primarily driven by inlet pressure/velocity profiles~\cite{caro2012mechanics}. 
(d) Certain boundary conditions become implicitly satisfied during training through the physics-informed constraints and do not require explicit inclusion in the loss function. An example is pressure prediction in the cavity flow problem, where only velocity components on the boundaries are satisfied during training, while the pressure is retrieved as a latent variable.
(e) Considering all possible boundary conditions can lead to over-constrained systems that may not be handled properly even with loss-balancing schemes.  It may also increase the computational cost without necessarily improving solution quality. This is particularly important for adaptive weighting loss balancing schemes, where each additional loss term requires separate weighting criteria or gradient calculations.

\subsubsection{2D Lid-driven Cavity Flow}

The loss function for the cavity problem is formulated as:

{\small
\begin{align}    
\label{cavity_loss}
    \mathcal{L}(\theta) &= \lambda_{\text{phy}}\|\mathcal{L}_{\text{phy}}\|_{\Omega} + \lambda_{\text{left}}\|\mathcal{L}_{\text{left}}\|_{\Gamma_{\text{left}}} + \lambda_{\text{right}}\|\mathcal{L}_{\text{right}}\|_{\Gamma_{\text{right}}} \notag \\
    &\quad + \lambda_{\text{bottom}}\|\mathcal{L}_{\text{bottom}}\|_{\Gamma_{\text{bottom}}} + \lambda_{\text{up}}\|\mathcal{L}_{\text{up}}\|_{\Gamma_{\text{up}}} \notag \\
    &\quad + \lambda_{\text{initial}}\|\mathcal{L}_{\text{initial}}\|_{\Omega_0} 
\end{align}
}
where $\mathcal{L}_{\text{phy}}(\theta) = \mathcal{L}_{\text{continuity}} + \mathcal{L}_{f_u} + \mathcal{L}_{f_v}$ represents the Navier-Stokes residuals such that:

{\small
\begin{align}
    \mathcal{L}_{r_u}(\theta) &= \text{MSE} \left[(\frac{\partial \hat{u}}{\partial t} + \hat{u}  \frac{\partial \hat{u}}{\partial x} + \hat{v} \frac{\partial \hat{u}}{\partial y}) + \frac{1.0}{ \rho}  \frac{\partial \hat{p}}{\partial x} - \nu  (\frac{\partial^2 \hat{u}}{\partial x^2} + \frac{\partial^2 \hat{u}}{\partial y^2})\right] \label{eq:ns_loss}\\
    \mathcal{L}_{r_v}(\theta) &=\text{MSE}\left[(\frac{\partial \hat{v}}{\partial t} + (\hat{u}  \frac{\partial \hat{v}}{\partial t} + \hat{v}  \frac{\partial \hat{v}}{\partial y} + \frac{1.0}{ \rho}  \frac{\partial \hat{p}}{\partial y} - \nu  (\frac{\partial^2 \hat{v}}{\partial x^2}+ \frac{\partial^2 \hat{v}}{\partial y^2})\right] \notag\\
    \mathcal{L}_{r_c}(\theta) &= \text{MSE}\left[(\frac{\partial \hat{u}}{\partial x} + \frac{\partial \hat{v}}{\partial y})\right] \notag
 \end{align}
}
and the boundary/initial conditions are:

{\small
\begin{align*}
    \mathcal{L}_{\text{bc}} &= \text{MSE}[(\hat{u} - u_{\text{bc}})^2 + (\hat{v} - v_{\text{bc}})^2] \\
    \mathcal{L}_{\text{up}} &= \text{MSE}[(\hat{u} - \vec{1.0})^2 + (\hat{v} - v_{\text{bc}})^2] \\
    \mathcal{L}_{\text{initial}} &= \text{MSE}[(\hat{u} - u_0)^2 + (\hat{v} - v_0)^2 + (\hat{p} - p_0)^2]
\end{align*}
}
\noindent where bc represents the corresponding left/right/bottom domain.

\subsubsection{Plane Poiseuille Flow}
We do not consider all boundary conditions in Eq.~\ref{eq:poiseuille1_loss}. The loss function for Poiseuille flow incorporates inlet boundary conditions and gradient constraints:

{\small
\begin{align*}    \label{poiseuille_loss}
    \mathcal{L}(\theta) &= \lambda_{\text{phy}}\|\mathcal{L}_{\text{phy}}\|_{\Omega} + \lambda_{\text{inlet}}\|\mathcal{L}_{\text{inlet}}\|_{\Gamma_{\text{inlet}}} + \lambda_{\text{initial}}\|\mathcal{L}_{\text{initial}}\|_{\Omega_0} \\
    &+ \lambda_{\text{wall}} \|\mathcal{L}_{\text{wall}}\|_{\Gamma_{\text{wall}}}
\end{align*}
}
\noindent with $\mathcal{L}_{\text{phy}}$ similar to Eq.~\ref{eq:ns_loss} with the corresponding viscosity and density values and specific boundary conditions:

{\small
\begin{align*}
    \mathcal{L}_{\text{inlet}} &= \text{MSE}[(\hat{u} - \vec{0.2})^2 + (\hat{p} - p_{\text{inlet}})^2 + (\frac{\partial \hat{p}}{\partial x})^2] \\
    \mathcal{L}_{\text{wall}} &= \text{MSE}[(\hat{p} - p_{\text{wall}})^2+ (\frac{\partial \hat{p}}{\partial y})^2] \\
    \mathcal{L}_{\text{initial}} &= \text{MSE}[(\hat{u} - \vec{0.2})^2 + \hat{v}^2 + \hat{p}^2] \\
\end{align*}
}

\subsubsection{2D Blood-flow in a tube with slip walls (BFS-Slip)}

The loss function for blood flow with slip boundary conditions is:

{\small
\begin{align*}    
    \mathcal{L}(\theta) &= \lambda_{\text{phy}}\|\mathcal{L}_{\text{phy}}\|_{\Omega} + \lambda_{\text{inlet}}\|\mathcal{L}_{\text{inlet}}\|_{\Gamma_{\text{inlet}}} \\
    &+ \lambda_{\text{outlet}}\|\mathcal{L}_{\text{outlet}}\|_{\Gamma_{\text{outlet}}}  + \lambda_{\text{wall}}\|\mathcal{L}_{\text{wall}}\|_{\Gamma_{\text{wall}}} \notag \\
    &+ \lambda_{\text{initial}}\|\mathcal{L}_{\text{initial}}\|_{\Omega_0} \notag
\end{align*}
}
where

{\small
\begin{align*}
    \mathcal{L}_{\text{inlet}} &= \text{MSE}[(\hat{u} - \vec{0.2})^2 + \hat{v}^2] \\
    \mathcal{L}_{\text{outlet}} &= \text{MSE}[(\hat{u} - u_{\text{outlet}})^2 + (\hat{v} - v_{\text{outlet}})^2  + \hat{p}^2 + (\frac{\partial \hat{u}}{\partial x})^2 + (\frac{\partial \hat{v}}{\partial x})^2] \\
    \mathcal{L}_{\text{initial}} &= \text{MSE}[(\hat{u} - \vec{0.2})^2 + \hat{v}^2 + \hat{p}^2] \\
\end{align*}
}
The physical loss $\mathcal{L}_{phy}$ is similar to Eq.~\ref{eq:ns_loss}  with the corresponding viscosity and density values.

\subsubsection{2D Blood-flow in a Tube with No-Slip Walls (BFS-No-Slip)}

Blood flow simulation using a no-slip case focuses on essential boundary conditions with a simplified formulation as follows:

{\small
\begin{align*}
    \mathcal{L}(\theta) &= \lambda_{\text{phy}}\|\mathcal{L}_{\text{phy}}\|_{\Omega} + \lambda_{\text{inlet}}\|\mathcal{L}_{\text{inlet}}\|_{\Gamma_{\text{inlet}}} \\
    &+ \lambda_{\text{initial}}\|\mathcal{L}_{\text{initial}}\|_{\Omega_0} \notag
\end{align*}
}
\noindent with:

{\small
\begin{align*}
    \mathcal{L}_{\text{inlet}} &= \text{MSE}[(\hat{u} -  \vec{0.2})^2 + \hat{v}^2 + (\hat{p} - p_{\text{inlet}})^2 + (\frac{\partial \hat{p}}{\partial x})^2] \\
    \mathcal{L}_{\text{wall}} &= \text{MSE}[(\hat{p} - p_{\text{wall}})^2+ (\frac{\partial \hat{p}}{\partial y})^2] \\
    \mathcal{L}_{\text{initial}} &= \text{MSE}[(\hat{u} - \vec{0.2})^2 + \hat{v}^2 + \hat{p}^2] \\
\end{align*}
}
The physical loss $\mathcal{L}_{\text{phy}}$ is similar to Eq.~\ref{eq:ns_loss} with the corresponding viscosity and density values. 

Our initial experiments showed that explicitly imposing the no-slip boundary condition on the loss function degrades overall training quality, whereas implicit satisfaction through physics constraints and other boundary conditions yields better convergence and more accurate velocity profiles. Therefore, in our loss function design, we do not explicitly enforce the no-slip boundary on the walls in this case.

\subsection{Computational Complexity}

\begin{table}[t]
\centering 
\ssmall
\caption{
Complexity of loss balancing with various activations, where $O$: asymptotic notation; $n$: units per hidden layer; $L$: number of hidden layers; $k$: grid size; $d$: spline order; $B$: batch size; and $m$: number of PINN loss terms.
}
\label{tab:basis_functions}
\begin{tabular}{@{}
                p{0.10\textwidth}
                p{0.14\textwidth}
                p{0.12\textwidth}
                @{}
                }
    \toprule
    \textbf{Weighting Method} & \textbf{B-spline + SiLU} &  \textbf{Tanh} \\ 
    \midrule
    Heuristic Search & $\mathcal{O}(n^2 \cdot L \cdot (k + d))$ & $\mathcal{O}(n^2 \cdot L)$\\ 
    RBA  & $\mathcal{O}(n^2 \cdot L \cdot (k + d) + B \cdot m)$ & $\mathcal{O}(n^2 \cdot L + B \cdot m)$\\ 
    LRA  & $\mathcal{O}(n^2 \cdot L \cdot (k + d) + m)$ & $\mathcal{O}(n^2 \cdot L + m)$\\ 
    SA  & $\mathcal{O}(n^2 \cdot L \cdot (k + d) + B \cdot m)$ & $\mathcal{O}(n^2 \cdot L + B \cdot m)$\\ 
    Grad-Norm  & $\mathcal{O}(n^2 \cdot L \cdot (k + d) + m)$ & $\mathcal{O}(n^2 \cdot L + m)$\\
    \bottomrule
    \end{tabular}
\end{table}

\red{
Table~\ref{tab:basis_functions} presents the computational complexity of loss balancing schemes using the Tanh and B-spline+SiLU, where $O$: asymptotic notation; $n$: units per hidden layer (assuming uniform number of units for all hidden payers); $L$: number of hidden layers; $k$: grid size; $d$: degree of polynomial basis (spline order); $B$: batch size; and $m$: number of PINN loss terms.
Using B-spline+SiLU is inherently more computationally expensive than using Tanh. This is because it adds a $(k + d)$ factor, related to its grid size and polynomial degree, to the standard fully connected network training cost ($\mathcal{O}(n^2 \cdot L)$). 
Moreover, adaptive loss balancing schemes, such as RBA and SA introduce a significant overhead of $\mathcal{O}(B \cdot m)$. This overhead arises from their point-wise strategy, which requires calculating a separate weight for each sample within a batch. 
In contrast, LRA and GradNorm are more efficient, adding a smaller overhead of only $\mathcal{O}(m)$. They achieve this by using a single scalar weight for each loss term, although this requires an extra gradient computation step.
}

\subsection{Implementation and Experiment Setup}

We configured and trained all models under similar conditions using the Adam optimizer, with an initial learning rate of 0.001. The training process spans 60,000 epochs, during which loss weights are updated at the end of each epoch when loss balancing schemes are applied. For network initialization, we utilize Xavier normal initialization for all linear layers, while biases are initialized to zero. 
The B-spline networks include additional parameters: a grid size of 5, a spline order of 3, and noise scale parameters set to 0.1. 
For the tanh activation function, we employed a network architecture of [3, 100, 100, 100, 100, 3] for all cases and adaptive schemes. In contrast, for the B-spline+SiLU, we used a network architecture of [3, 50, 50, 50, 3]. All code implementations are carried out using the PyTorch library. 

To reduce computation and ensure a well-distributed and representative training dataset, we first generate collocation points using the Sobol sequence, which provides low-discrepancy quasi-random samples that achieve better space-filling properties compared to purely random sampling. During training, we further enhance diversity and reduce sampling bias by applying random mini-batch sampling with a batch size of 128 drawn from the Sobol-generated dataset at each iteration.

To assess the models' performance, we employ Root Mean Square Error (RMSE) on randomly selected test datasets from the numerical solutions as $\text{RMSE} = \sqrt{\frac{1}{n} \sum_{i=1}^{n} (f_i - \hat{f}_i)^2}$, 
%
% \begin{equation}
% \text{RMSE} = \sqrt{\frac{1}{n} \sum_{i=1}^{n} (f_i - \hat{f}_i)^2}
% \end{equation}
%
\noindent where $f_i$, $\hat{f}_i$, and $n$ represent reference solution, PINN predictions, and total number of evaluation points, respectively. 
We prefer using the RMSE over relative error metrics, such as the relative mean square error, because the latter can be particularly sensitive when dealing with flow regions that have very small magnitudes or near-zero values. This sensitivity is evident in the v-velocity values observed in certain BFS cases, as illustrated in the contour plot errors shown in Fig.~\ref{fig:poiseulli_fixed_bspline},~\ref{fig:slip_sa_bspline},~\ref{fig:noslip_sa_bspline}. 

\red{
To evaluate the computational cost of each adaptive weighting method in terms of both execution time and GPU memory usage during training,  we record the wall-clock duration for each iteration. Memory usage is assessed from the PyTorch profiler. The time and memory usage were averaged across all iterations, which included the forward pass, loss computation, backward propagation, and optimizer step for each iteration.
}
\red{
Finally, measurements are conducted using experimental results from a personal laptop with Intel Core i7-10700 CPU, 32 GB DDR4 RAM, and an NVIDIA GeForce RTX 2060 GPU.
}

%% file: sections/5_results.tex
\section{Experimental Results}

%------------------------------------------------------------------------------%
% NOTE: This is part of result section, moved here to fix placement issue.
\input{sections/5_results_tables_figures}
%------------------------------------------------------------------------------%

\subsection{Performance Comparison Across Loss Balancing Schemes}
Table~\ref{tab:performance} shows the performance comparison of loss balancing schemes for PINN across four fluid flow test cases using two different activation functions. The Heuristic search here is based on the best empirical settings.
The results show significant performance variations both across different loss balancing schemes and activation function choices.

Our primary hypothesis that the effectiveness of multi-objective loss balancing depends on both the balancing strategy and the underlying activation function is strongly supported by the experimental results. Across all fluid flow scenarios, the choice of activation function consistently influences performance, with B-spline+SiLU activation generally outperforming fixed Tanh activation functions. 

Despite using approximately 2.5 times more parameters (53,001 vs. 20,904), the B-spline+SiLU networks consistently achieve better performance per parameter, suggesting better function approximation capabilities rather than simple over-parameterization. Table~\ref{tab:performance} quantifies these improvements, showing percentage reductions in RMSE when using B-spline+SiLU compared to Tanh activation. The improvements are substantial and consistent: Heuristic search shows 22.4-57.6\% improvement across cases, RBA demonstrates 17.8-41.4\% improvement, and LRA achieves up to 95.2\% improvement in the BFS-Slip case. The SA scheme shows more variable performance, with improvements ranging from 7.4\% to 67.9\%, and showing a 12.3\% degradation in the BFS-No-Slip case.

\subsection{Loss Balancing Strategy Effectiveness}
Across various loss balancing schemes considered, LRA demonstrates improvements with both types of activation. The Heuristic Search method demonstrates reliable performance, while RBA shows moderate effectiveness. SA performance is highly case-dependent, and GradNorm proves unreliable, particularly with learnable activations.

\subsection{Training Convergence Analysis}
Fig.~\ref{fig:loss_history} shows the training loss convergence for the loss balancing schemes using two different network architectures: Tanh activation (top row) and B-spline+SiLU activation (bottom row). The results indicate that networks employing B-spline+SiLU activations generally achieve faster convergence and lower final loss values compared to those using Tanh activations across all weighting schemes and flow problems. The LRA with B-spline+SiLU shows particularly effective convergence behavior for the Cavity flow problem. However, it exhibits more oscillatory behavior during optimization with the BFS problems. These fluctuations in the LRA highlight the method's high sensitivity to gradient changes, especially in BFS problems.

\subsection{Solution Quality Visualization}
Fig.~\ref{fig:cavity_fixed_bspline} shows the contour plots of the best PINN solutions at the end of the simulation for the Cavity problem using LRA weighting with B-spline+SiLU activation. The reference solution (left column) shows the flow pattern with maximum $u$-velocity of 9.7e-01 at the top boundary and a prominent central vortex. The PINN predictions (middle column) accurately capture the flow structure, with velocity magnitudes closely matching the reference solution. The pressure field shows the typical distribution with smooth gradients throughout the domain. The absolute error plots (right column) indicate excellent agreement, demonstrating the effectiveness of the LRA weighting approach for this canonical flow problem.

Fig.~\ref{fig:poiseulli_fixed_bspline} shows the contour plots of the numerical solution and the PINN solution for the Poiseulle problem using fixed heuristic weighting with B-spline+SiLU activation at the last time step of the simulation. The reference solution displays the characteristic parabolic velocity profile with a maximum $u$-velocity of 2.9e-01 at the channel centerline. The $v$-velocity component remains near zero throughout the domain, as expected for fully developed flow. The pressure field shows a linear decrease in the streamwise direction. The PINN predictions accurately reproduce these features, with the neural network successfully learning the analytical solution behavior. 

Fig.~\ref{fig:slip_sa_bspline}  shows the contour plots of the PINN solution for BFS-Slip using SA weighting with B-spline+SiLU activation at the end of the simulation. The reference solution shows the expected parabolic-like velocity profile with a maximum $u$-velocity of 2.0e-01, while maintaining near-zero $v$-velocity throughout most of the domain. The pressure field exhibits a smooth gradient along the flow direction. The PINN solution demonstrates excellent accuracy in capturing the flow physics, with the SA weighting method effectively balancing the multiple loss terms.

Fig.~\ref{fig:noslip_sa_bspline} shows the contour plots of the PINN solution for blood flow with no-slip boundary conditions discussed in section~\ref{sec:use-cases} using SA weighting with B-spline+SiLU activation. The reference solution exhibits the classical Poiseuille profile with a maximum $u$-velocity of 4.0e-01 and zero velocity at the walls. The $v$-velocity remains minimal throughout the domain, and the pressure shows a linear distribution. The PINN predictions accurately capture these characteristics, with the SA method demonstrating robust performance in handling the no-slip constraints.

\subsection{Computational Efficiency Analysis}

\red{
The computational efficiency analysis presented in Table~\ref{tab:computation-cost} reveals important trade-offs between accuracy and computational overhead when employing different activation functions and loss balancing schemes. 
As B-spline+SiLU networks require approximately 2.5 times more parameters (53,001 vs. 20,904), they exhibit increased execution time per iteration (ranging from 0.28-0.58 seconds compared to 0.04-0.11 seconds for Tanh), and the computational overhead varies significantly across different loss balancing methods. Among the adaptive strategies, RBA demonstrates the most efficient balance between computational cost and performance improvement, with execution times of 0.28-0.47 seconds and memory consumption ranging from 80.35-116.84 MB. 
LRA shows the highest computational overhead among successful methods, with execution times reaching 0.36-0.58 seconds and memory consumption up to 281.90 MB. The SA method maintains moderate computational efficiency with execution times of 0.30-0.38 seconds while achieving competitive accuracy improvements.
}

\red{
While Tanh-based networks are faster and more memory-efficient, their accuracy is generally lower compared to B-spline+SiLU networks.
The learnable B-spline+SiLU activation functions introduce additional parameters and spline computations at each layer, which significantly increase both runtime and memory usage.
Among the adaptive strategies, lightweight approaches such as RBA, which only require normalized residual updates, impose minimal extra cost, whereas gradient-intensive methods like LRA and GradNorm significantly increase both runtime and memory due to repeated gradient norm calculations and weight updates. 
Notably, the computational cost increase is partially justified by the substantial accuracy gains achieved, particularly in complex flow scenarios where traditional methods struggle with convergence.
}

%% file: sections/5_results_tables_figures.tex
\begin{table*}[t]
\centering
\ssmall
\caption{
Performance comparison of loss balancing schemes. The table presents the RMSE values for velocity components ($u$ and $v$) and pressure ($p$), along with the final training loss values (L\textsubscript{f}).
The row $\Delta$\textsubscript{L\textsubscript{f}} (\%)$\uparrow$ shows the relative percentage difference between Tanh and B-spline+SiLU, calculated as: $\frac{\text{RMSE}_{\text{Tanh}} - \text{RMSE}_{\text{B-spline}}}{\text{RMSE}_{\text{Tanh}}} \times 100\%$.
``F'' indicates failure when testing errors exceed 90\%.  
}
\label{tab:performance}
\begin{tabular}{
    @{}
    p{0.07\textwidth}@{\hspace{0.1pt}}
    p{0.07\textwidth}@{\hspace{0.1pt}}
    p{0.01\textwidth}@{\hspace{0.1pt}}
    p{0.065\textwidth}@{\hspace{0.1pt}}
    p{0.06\textwidth}@{\hspace{0.1pt}}
    p{0.06\textwidth}@{\hspace{0.1pt}}
    !{\vline}
    p{0.01\textwidth}@{\hspace{0.1pt}}
    p{0.065\textwidth}@{\hspace{0.1pt}}
    p{0.06\textwidth}@{\hspace{0.1pt}}
    p{0.06\textwidth}@{\hspace{0.1pt}}
    !{\vline}
    p{0.01\textwidth}@{\hspace{0.1pt}}
    p{0.065\textwidth}@{\hspace{0.1pt}}
    p{0.06\textwidth}@{\hspace{0.1pt}}
    p{0.06\textwidth}@{\hspace{0.1pt}}
    !{\vline}
    p{0.01\textwidth}@{\hspace{0.1pt}}
    p{0.065\textwidth}@{\hspace{0.1pt}}
    p{0.06\textwidth}@{\hspace{0.1pt}}
    p{0.06\textwidth}@{\hspace{0.1pt}}
    @{}}
    
    \toprule
    
    \multirow{2}{*}{\textbf{Scheme}}& 
    \multirow{2}{*}{\textbf{Activation}}&
    \multicolumn{4}{c}{\textbf{Cavity Flow}}& 
    \multicolumn{4}{c}{\textbf{Poiseuille}}& 
    \multicolumn{4}{c}{\textbf{BFS-Slip}}& 
    \multicolumn{4}{c}{\textbf{BFS-No-Slip}}\\
    
    \cmidrule(){3-18}& & 
    
    &
    \textbf{RMSE}$\downarrow$& 
    \textbf{L\textsubscript{f}}$\downarrow$&
    \textbf{$\Delta$\textsubscript{L\textsubscript{f}} (\%)$\uparrow$}&
    &
    \textbf{RMSE}$\downarrow$& 
    \textbf{L\textsubscript{f}}$\downarrow$&
    \textbf{$\Delta$\textsubscript{L\textsubscript{f}} (\%)$\uparrow$}&
    &
    \textbf{RMSE}$\downarrow$& 
    \textbf{L\textsubscript{f}}$\downarrow$&
    \textbf{$\Delta$\textsubscript{L\textsubscript{f}} (\%)$\uparrow$}&
    &
    \textbf{RMSE}$\downarrow$& 
    \textbf{L\textsubscript{f}}$\downarrow$&
    \textbf{$\Delta$\textsubscript{L\textsubscript{f}} (\%)$\uparrow$} \\
    
    \cmidrule(){1-18}

    % ---------------- Heuristic Search ---------------- %
    \multirow{7}{*}{\makecell[l]{Heuristic\\Search}}& \multirow{3}{*}{Tanh}& $u$& 5.052e-02& \multirow{3}{*}{2.34e-02}& \multirow{6}{*}{31.1}& $u$& 9.349e-03& \multirow{3}{*}{1.16e-04}& \multirow{6}{*}{57.6}& $u$& 2.202e-05& \multirow{3}{*}{5.13e-09}& \multirow{6}{*}{22.4}& $u$& 1.218e-02& \multirow{3}{*}{2.34e-05}& \multirow{6}{*}{23.5}\\
    
    & & $v$& 6.436e-02& & & $v$& 1.024e-03& & & $v$& 8.379e-06& & & $v$& 1.337e-03& & \\
    & & $p$& 4.693e-02& & & $p$& 2.275e-03& & & $p$& 1.608e-05& & & $p$& 2.596e-03& & \\

    \noalign{\vskip 1ex}
    
    & \multirow{3}{*}{\makecell[l]{B-spline\\+\\SiLU}}& 
    $u$& 2.942e-02& \multirow{3}{*}{9.13e-03}& & $u$& 4.860e-03& \multirow{3}{*}{5.10e-05}& & $u$& 3.360e-05& \multirow{3}{*}{3.48e-09}& & $u$& 1.384e-02& \multirow{3}{*}{1.72e-06}& \\
    & & $v$& 4.480e-02& & & $v$& 3.988e-04& & & $v$& 2.377e-06 & & & $v$& 6.288e-04& & \\
    & & $p$& 3.419e-02& & & $p$& 8.259e-04& & & $p$& 8.330e-06 & & & $p$& 1.786e-03& & \\

    \midrule
    
    % ---------------- RBA ---------------- %
    \multirow{7}{*}{RBA} 
    & \multirow{3}{*}{Tanh}& $u$& 8.292e-02& \multirow{3}{*}{7.31e-02}& \multirow{6}{*}{20.1}& $u$& 1.013e-02& \multirow{3}{*}{2.71e-04}& \multirow{6}{*}{21.6}& $u$& 4.949e-05& \multirow{3}{*}{4.35e-09}& \multirow{6}{*}{17.8}& $u$& 3.570e-02& \multirow{3}{*}{2.32e-04}& \multirow{6}{*}{41.4}\\
    
    & & $v$& 1.045e-01& & & $v$& 5.951e-04& & & $v$& 5.193e-06& & & $v$& 8.306e-04& & \\
    & & $p$& 6.361e-02& & & $p$& 2.332e-03& & & $p$& 1.755e-05& & & $p$& 1.089e-02& & \\

    \noalign{\vskip 1ex}

    & \multirow{3}{*}{\makecell[l]{B-spline\\+\\SiLU}}& $u$& 7.545e-02& \multirow{3}{*}{3.51e-02}& & $u$& 8.828e-03& \multirow{3}{*}{2.48e-04}& & $u$& 4.586e-05& \multirow{3}{*}{4.28e-10}& & $u$& 2.455e-02& \multirow{3}{*}{4.19e-06}& \\
    & & $v$& 8.171e-02& & & $v$& 5.502e-04& & & $v$& 4.455e-06 & & & $v$& 6.894e-04& & \\
    & & $p$& 4.477e-02& & & $p$& 1.296e-03& & & $p$& 1.198e-05 & & & $p$& 2.618e-03& & \\

    \midrule

    % ---------------- LRA ---------------- %
    \multirow{7}{*}{LRA} 
    & \multirow{3}{*}{Tanh}& $u$& 3.061e-02& \multirow{3}{*}{1.04e-01}& \multirow{6}{*}{39.9}& $u$& 1.304e-02& \multirow{3}{*}{5.30e-04}& \multirow{6}{*}{59.1}& $u$& 7.886e-04& \multirow{3}{*}{7.49e-06}& \multirow{6}{*}{95.2}& $u$& 1.426e-02& \multirow{3}{*}{5.26e-05}& \multirow{6}{*}{-12.3}\\
    
    & & $v$& 3.668e-02& & & $v$& 2.329e-03& & & $v$& 1.022e-03& & & $v$& 4.629e-04& & \\
    & & $p$& 3.979e-02& & & $p$& 5.076e-03& & & $p$& 7.337e-04& & & $p$& 3.159e-03& & \\

    \noalign{\vskip 1ex}

    & \multirow{3}{*}{\makecell[l]{B-spline\\+\\SiLU}}& $u$& 2.500e-02& \multirow{3}{*}{8.58e-04}& & $u$& 9.218e-03& \multirow{3}{*}{1.75e-04}& & $u$& 1.019e-04& \multirow{3}{*}{2.23e-07}& & $u$& 2.371e-02& \multirow{3}{*}{4.69e-07}& \\
    & & $v$& 1.356e-02& & & $v$& 4.864e-04& & & $v$& 1.763e-06 & & & $v$& 4.956e-04& & \\
    & & $p$& 2.448e-02& & & $p$& 1.577e-03& & & $p$& 9.513e-06 & & & $p$& 2.010e-03& & \\

    \midrule

    % ---------------- SA ---------------- %
    \multirow{7}{*}{SA} 
    & \multirow{3}{*}{Tanh}& $u$& 5.932e-02& \multirow{3}{*}{2.77e-02}& \multirow{6}{*}{7.4}& $u$& 9.535e-03& \multirow{3}{*}{7.26e-05}& \multirow{6}{*}{19.6}& $u$& 4.355e-05& \multirow{3}{*}{1.06e-08}& \multirow{6}{*}{67.9}& $u$& 1.152e-02& \multirow{3}{*}{2.27e-05}& \multirow{6}{*}{27.3}\\
    
    & & $v$& 8.179e-02& & & $v$& 7.511e-04& & & $v$& 1.946e-05& & & $v$& 1.569e-03& & \\
    & & $p$& 4.663e-02& & & $p$& 1.774e-03& & & $p$& 5.609e-05& & & $p$& 2.830e-03& & \\

    \noalign{\vskip 1ex}

    & \multirow{3}{*}{\makecell[l]{B-spline\\+\\SiLU}}& $u$& 5.839e-02& \multirow{3}{*}{3.90e-02}& & $u$& 1.236e-02& \multirow{3}{*}{1.04e-04}& & $u$& 2.160e-05& \multirow{3}{*}{1.57e-09}& & $u$& 1.565e-02& \multirow{3}{*}{1.93e-06}& \\
    & & $v$& 8.404e-02& & & $v$& 4.520e-04& & & $v$& 5.377e-06 & & & $v$& 4.583e-04& & \\
    & & $p$& 3.574e-02& & & $p$& 9.117e-04& & & $p$& 1.072e-05 & & & $p$& 1.502e-03& & \\

    \midrule

    % ---------------- GradNorm ---------------- %
    \multirow{7}{*}{GradNorm} 
    & \multirow{3}{*}{Tanh}& $u$& 2.057e-01& \multirow{3}{*}{1.05e+00}& \multirow{6}{*}{}& $u$& &  \multirow{3}{*}{\makecell[l]{F}}& \multirow{6}{*}{}& $u$& 4.501e-05& \multirow{3}{*}{5.62e-09}& \multirow{6}{*}{}& $u$& &  \multirow{3}{*}{\makecell[l]{F}}& \multirow{6}{*}{}\\
    
    & & $v$& 1.294e-01& & & $v$& & & & $v$& 1.135e-05& & & $v$& & & \\
    & & $p$& 8.070e-02& & & $p$& & & & $p$& 5.851e-06& & & $p$& & & \\

    \noalign{\vskip 1ex}

    & \multirow{3}{*}{\makecell[l]{B-spline\\+\\SiLU}}& $u$& &  \multirow{3}{*}{\makecell[l]{F}}& & $u$&  &  \multirow{3}{*}{\makecell[l]{F}}& & $u$& & \multirow{3}{*}{\makecell[l]{F}}& & $u$& &  \multirow{3}{*}{\makecell[l]{F}}& \\
    & & $v$& & & & $v$& & & & $v$& & & & $v$& & & \\
    & & $p$& & & & $p$& & & & $p$& & & & $p$& & & \\

    \bottomrule
\end{tabular}
\end{table*}

\begin{table*}[t]
\centering
\ssmall
\caption{
\red{
Computational cost of the loss balancing schemes. T: average time per iteration (in seconds), M: average peak memory consumption per iteration (in megabytes), P: total number of trainable parameters, $\Delta$T and $\Delta$M are the difference between Tanh and B-spline+SiLU. Values for GradNorm are not captured as their testing errors exceed 90\%, see Table~\ref{tab:performance}.
} 
% \green{TODO: Need to make sure of these values again!}
}
\label{tab:computation-cost}
\begin{tabular}{
    @{}
    p{0.06\textwidth}@{\hspace{0.6pt}}
    p{0.08\textwidth}@{\hspace{0.8pt}}
    p{0.04\textwidth}@{\hspace{0.4pt}}
    p{0.01\textwidth}@{\hspace{0.1pt}}
    p{0.035\textwidth}@{\hspace{0.4pt}}
    p{0.035\textwidth}@{\hspace{0.4pt}}
    p{0.04\textwidth}@{\hspace{0.1pt}}
    p{0.04\textwidth}@{\hspace{0.4pt}}
    !{\vline}
    p{0.04\textwidth}@{\hspace{0.4pt}}
    p{0.01\textwidth}@{\hspace{0.1pt}}
    p{0.035\textwidth}@{\hspace{0.4pt}}
    p{0.035\textwidth}@{\hspace{0.4pt}}
    p{0.04\textwidth}@{\hspace{0.1pt}}
    p{0.04\textwidth}@{\hspace{0.4pt}}
    !{\vline}
    p{0.04\textwidth}@{\hspace{0.4pt}}
    p{0.01\textwidth}@{\hspace{0.1pt}}
    p{0.035\textwidth}@{\hspace{0.4pt}}
    p{0.035\textwidth}@{\hspace{0.4pt}}
    p{0.04\textwidth}@{\hspace{0.1pt}}
    p{0.04\textwidth}@{\hspace{0.4pt}}
    !{\vline}
    p{0.04\textwidth}@{\hspace{0.4pt}}
    p{0.01\textwidth}@{\hspace{0.1pt}}
    p{0.035\textwidth}@{\hspace{0.4pt}}
    p{0.035\textwidth}@{\hspace{0.4pt}}
    p{0.04\textwidth}@{\hspace{0.1pt}}
    p{0.04\textwidth}@{\hspace{0.4pt}}
    @{}}

    \toprule

    \multirow{2}{*}{\textbf{Scheme}}&
    \multirow{2}{*}{\textbf{Activation}}&
    \multicolumn{6}{c}{\textbf{Cavity Flow}}&
    \multicolumn{6}{c}{\textbf{Poiseuille}}&
    \multicolumn{6}{c}{\textbf{BFS-Slip}}&
    \multicolumn{6}{c}{\textbf{BFS-No-Slip}}\\

    \cmidrule(){3-26}&
    &
    \textbf{P}&
    &
    \textbf{T}$\downarrow$&
    \textbf{$\Delta$\textsubscript{T}}$\downarrow$&
    \textbf{M}$\downarrow$&
    \textbf{$\Delta$\textsubscript{M}}$\downarrow$&
    \textbf{P}&
    &
    \textbf{T}$\downarrow$&
    \textbf{$\Delta$\textsubscript{T}}$\downarrow$&
    \textbf{M}$\downarrow$&
    \textbf{$\Delta$\textsubscript{M}}$\downarrow$&
    \textbf{P}&
    &
    \textbf{T}$\downarrow$&
    \textbf{$\Delta$\textsubscript{T}}$\downarrow$&
    \textbf{M}$\downarrow$&
    \textbf{$\Delta$\textsubscript{M}}$\downarrow$&
    \textbf{P}&
    &
    \textbf{T}$\downarrow$&
    \textbf{$\Delta$\textsubscript{T}}$\downarrow$&
    \textbf{M}$\downarrow$&
    \textbf{$\Delta$\textsubscript{M}}$\downarrow$\\
    \cmidrule(){1-26}

 \multirow{2}{*}{\makecell[l]{Heuristic\\Search}}&  Tanh & 20904 & & 0.05 & \multirow{2}{*}{0.23} & 21.12 & \multirow{2}{*}{74.34} & 20904 & & 0.05 & \multirow{2}{*}{0.44} &  141.92 & \multirow{2}{*}{58.15} & 20904 & & 0.04 & \multirow{2}{*}{0.29} & 21.32 & \multirow{2}{*}{74.03} & 20904 & & 0.04 & \multirow{2}{*}{0.42} & 22.27 & \multirow{2}{*}{59.61} \\

 & B-spline+SiLU & 53001 & & 0.28 & &  95.46 & & 53001 & & 0.49 & &  200.07 & & 53001 & & 0.33 & & 95.35 & & 53001 & & 0.46 & & 81.88 & \\
 \cmidrule(){1-26}

 \multirow{2}{*}{RBA}&  Tanh & 20904 & & 0.04 & \multirow{2}{*}{0.24} & 41.77 & \multirow{2}{*}{75.07} & 20904 & & 0.04 & \multirow{2}{*}{0.24} & 119.73  & \multirow{2}{*}{59.90} & 20904 & & 0.04 & \multirow{2}{*}{0.33} & 30.27 & \multirow{2}{*}{68.95} & 20904 & & 0.06 & \multirow{2}{*}{0.41} & 23.82 & \multirow{2}{*}{56.53} \\

 & B-spline+SiLU & 53001 & & 0.28 & & 116.84 & & 53001 & & 0.28 & & 179.63 & & 53001 & & 0.37 & & 99.22 & & 53001 & & 0.47 & & 80.35 & \\

 \cmidrule(){1-26}

 \multirow{2}{*}{LRA}&  Tanh & 20904 & & 0.06 & \multirow{2}{*}{0.41} & 48.49 & \multirow{2}{*}{127.35} & 20904 & & 0.11 & \multirow{2}{*}{0.36} & 146.76 & \multirow{2}{*}{135.14} & 20904 & & 0.07 & \multirow{2}{*}{0.48} & 75.34 & \multirow{2}{*}{153.20} & 20904 & & 0.09 & \multirow{2}{*}{0.49} & 28.42 & \multirow{2}{*}{133.00} \\

 & B-spline+SiLU & 53001 & & 0.47 & &  175.84 & & 53001 & & 0.47 & & 281.90 & & 53001 & & 0.55 & & 228.54 & & 53001 & & 0.58 & & 161.42 & \\
 \cmidrule(){1-26}

 \multirow{2}{*}{SA}&  Tanh & 21800 & & 0.04 & \multirow{2}{*}{0.26} & 24.40 & \multirow{2}{*}{74.91} & 20904 & & 0.04 & \multirow{2}{*}{0.27} & 209.64 & \multirow{2}{*}{63.02} & 21672 & & 0.05 & \multirow{2}{*}{0.33} & 128.39 & \multirow{2}{*}{75.02} & 21288 & & 0.05 & \multirow{2}{*}{0.27} & 150.84 & \multirow{2}{*}{59.63} \\

 & B-spline+SiLU & 53897 & & 0.30 & & 99.31 & & 53001 & & 0.31 & & 272.66 & & 53769 & & 0.38 & & 203.41 & & 53385 & & 0.32 & & 210.47 & \\
 \cmidrule(){1-26}

 \multirow{2}{*}{GradNorm}&  Tanh & 20911 & & 0.08 & \multirow{2}{*}{} & 31.79 & \multirow{2}{*}{} & 20904 & & & \multirow{2}{*}{} & & \multirow{2}{*}{} & 20910 & & 0.09 & \multirow{2}{*}{} & 184.52 & \multirow{2}{*}{} & 20908 & & & \multirow{2}{*}{} & & \multirow{2}{*}{} \\

 & B-spline+SiLU & 53008 & & & & & & 53001 & & & & & & 53007 & & & & & & 53005 & & & & & \\
 \bottomrule
\end{tabular}
\end{table*}

\begin{figure*}[t]
\centering
\scriptsize
\begin{tabular}{cccc}
    \footnotesize{Cavity Flow} & \footnotesize{Poiseulle} & \footnotesize{ BFS-Slip} & \footnotesize{ BFS-No-Slip}\\

    \includegraphics[width=0.47\columnwidth, trim={0cm .10cm 0cm 0cm}, clip]{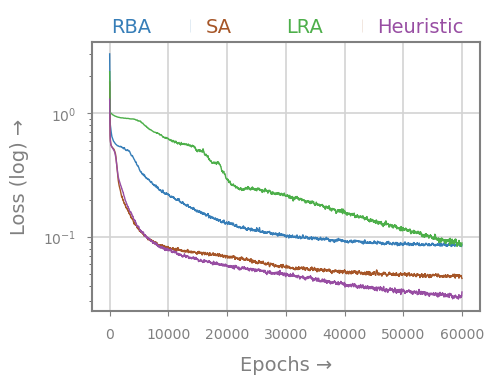} &
    \includegraphics[width=0.47\columnwidth, trim={0cm .10cm 0cm 0cm}, clip]{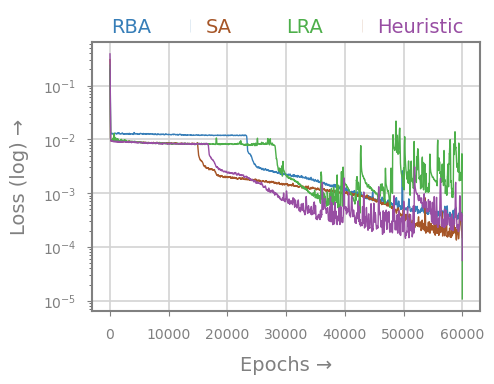} &
    \includegraphics[width=0.47\columnwidth, trim={0cm .10cm 0cm 0cm}, clip]{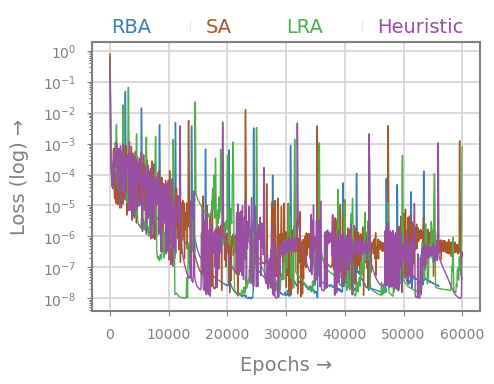} &

    \includegraphics[width=0.47\columnwidth, trim={0cm .10cm 0cm 0cm}, clip]{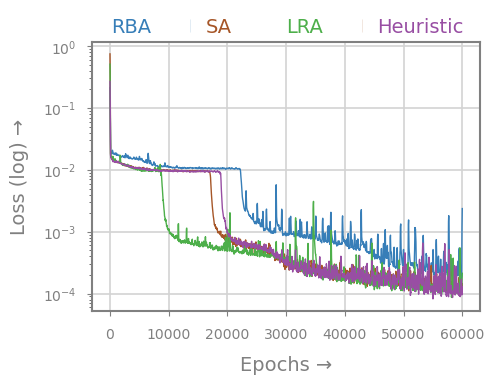} \\
    \multicolumn{4}{c}{ $-$ Tanh $-$}
    \\
    \includegraphics[width=0.47\columnwidth, trim={0cm .10cm 0cm 0cm}, clip]{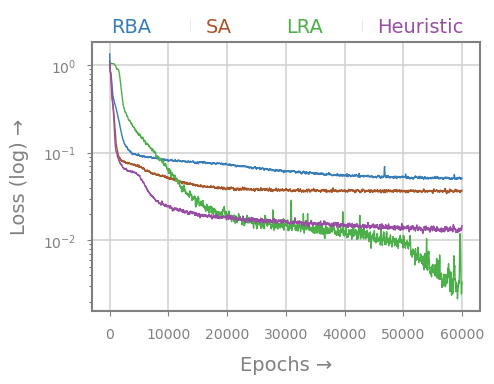} &
    \includegraphics[width=0.47\columnwidth, trim={0cm .10cm 0cm 0cm}, clip]{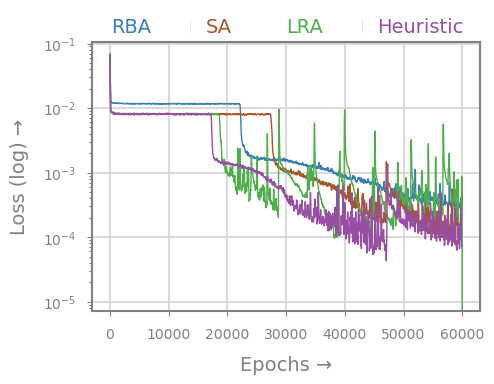} &
    \includegraphics[width=0.47\columnwidth, trim={0cm .10cm 0cm 0cm}, clip]{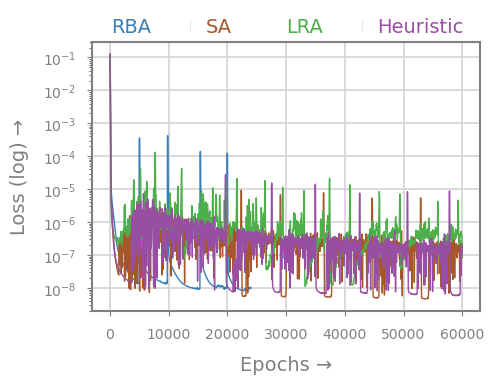} &
    \includegraphics[width=0.47\columnwidth, trim={0cm .10cm 0cm 0cm}, clip]{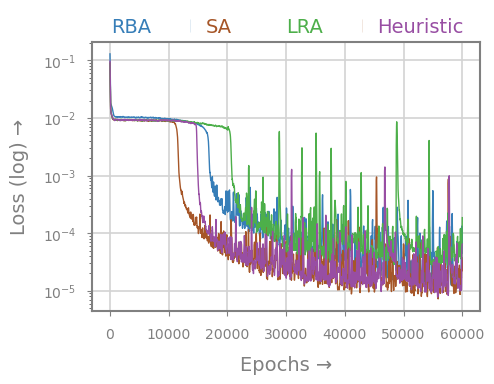} \\
    \multicolumn{4}{c}{$-$ B-spline+SiLU $-$}
\end{tabular}

\caption{ Training loss convergence curves for the loss balancing schemes across the selected fluid flow problems using Tanh (top) and B-spline+SiLU (bottom) activation functions.}
\label{fig:loss_history}
\end{figure*}

\begin{figure}[t]
\centering
    \includegraphics[width=0.9\columnwidth, trim={0cm 0cm 0cm 0cm}, clip]{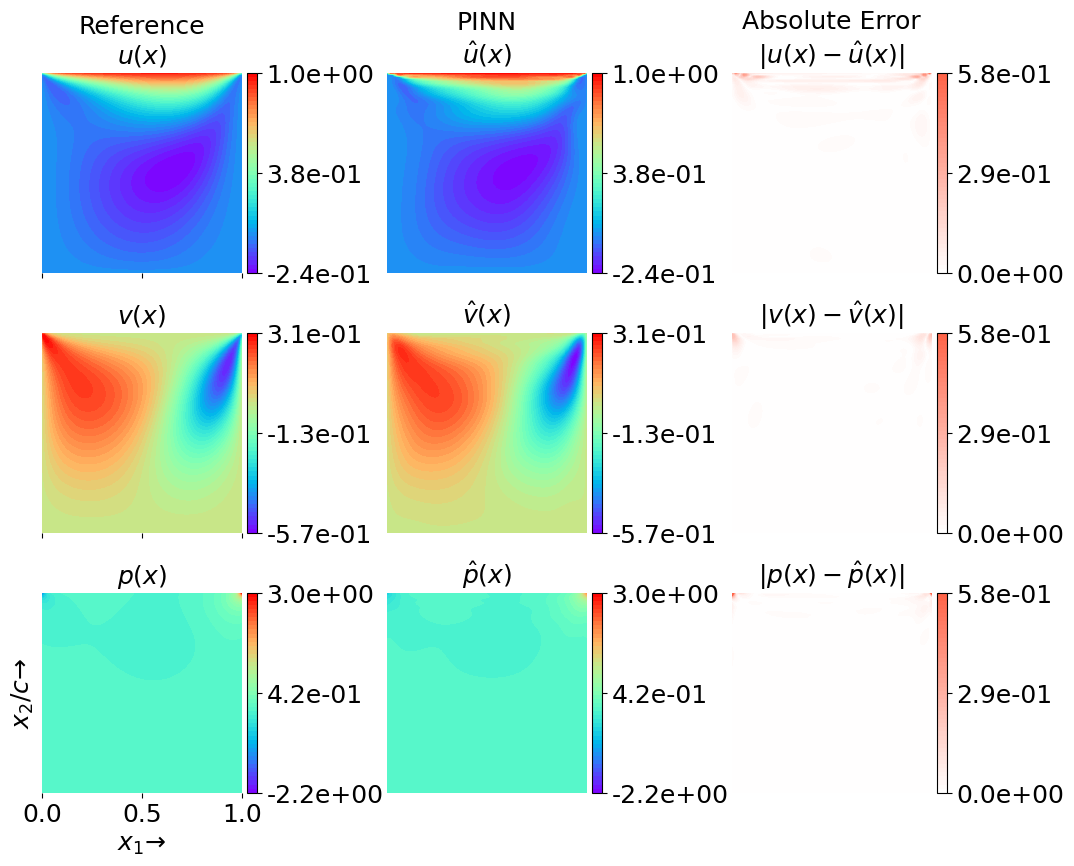} 
    \caption{Contour plots for the solution of the Cavity problem, showing velocity components ($u_x$ and $v_y$) and pressure ($p$) values at the end of the simulation (steady state). The PINN model yields the best results with LRA, as shown in Table~\ref{tab:performance}.
    }
    \label{fig:cavity_fixed_bspline}
\end{figure} 

\begin{figure}[t]
\centering
    \includegraphics[width=1.0\columnwidth, trim={0cm 0cm 0cm 0cm}, clip]{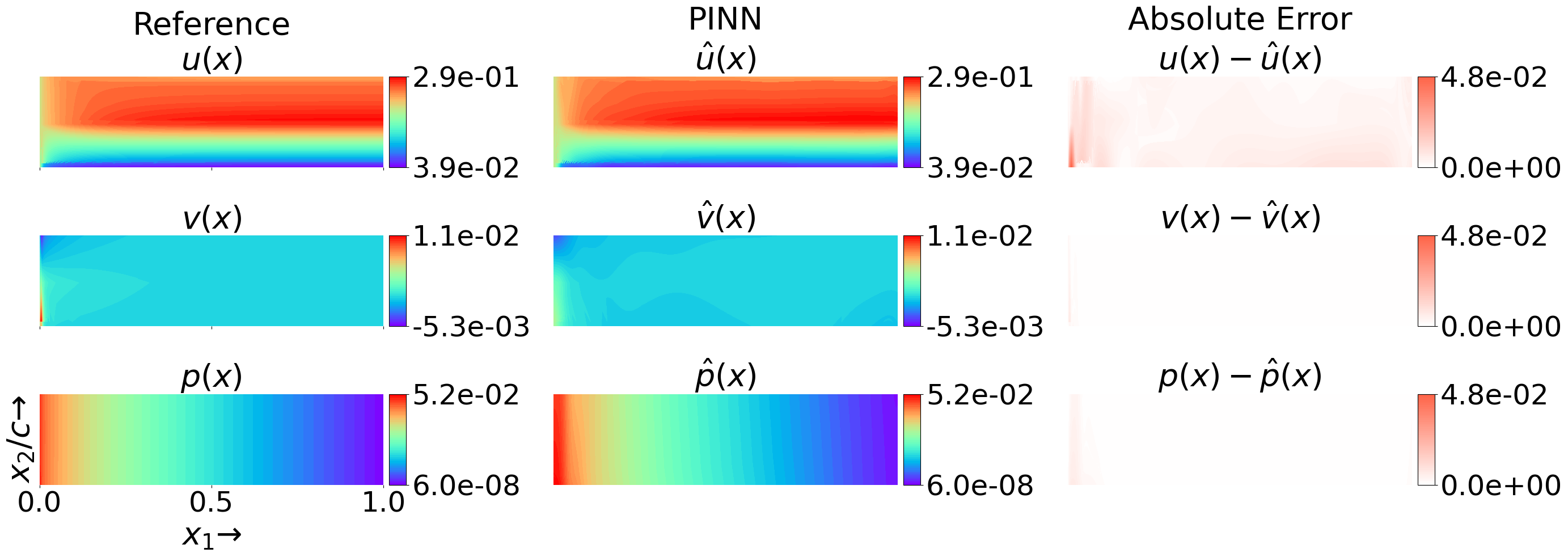} 
    \caption{Contour plots for the solution of the Poiseuille problem, showing velocity components ($u_x$ and $v_y$) and pressure ($p$) values at the steady state. The PINN model shows results with the fixed heuristic weighting with 0.1, 2, 2 for $\lambda_{phy}$, $\lambda_{bc}$, and $\lambda_{ic}$, respectively, which gives the best results as shown in Table~\ref{tab:performance}. }
    \label{fig:poiseulli_fixed_bspline}
\end{figure}

\begin{figure}[t]
\centering
    \includegraphics[width=1.0\columnwidth , trim={0cm 0cm 0cm 0cm}, clip]{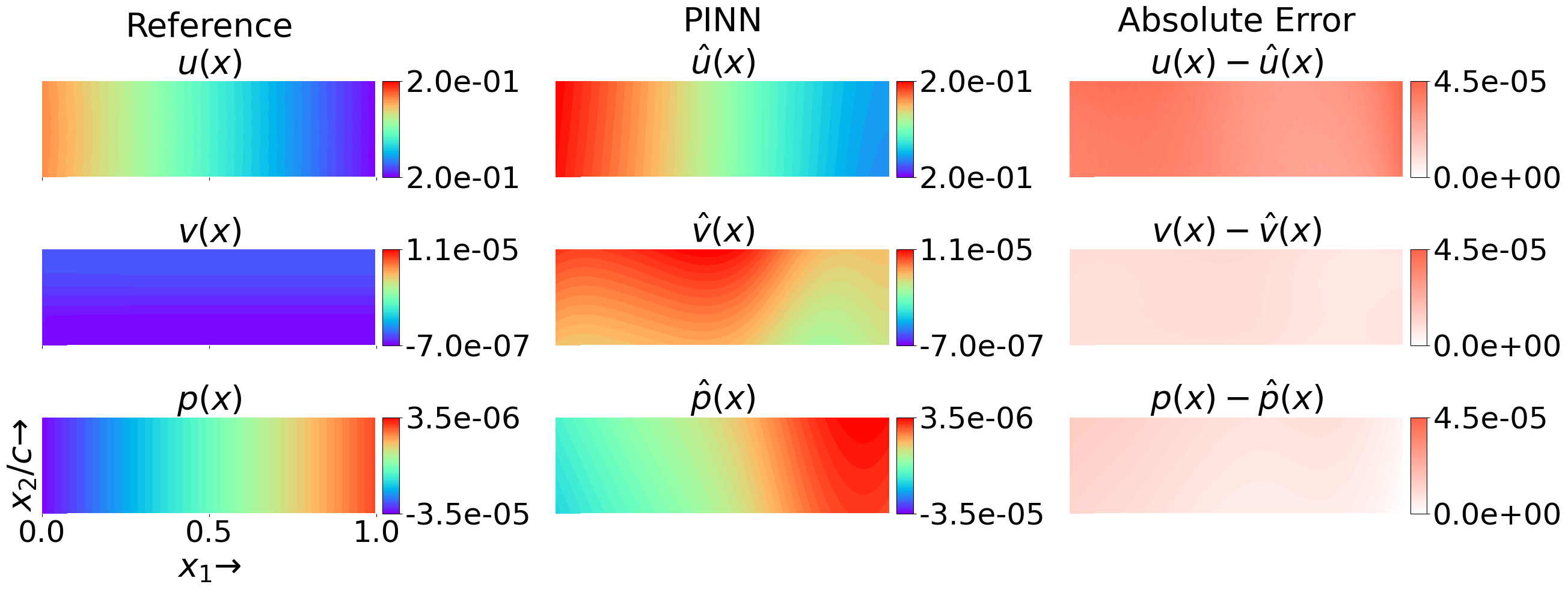} 
    \caption{\red{Contour plots for the solution of the BFS-Slip problem, showing velocity components ($u_x$ and $v_y$) and pressure ($p$) values at the steady state. PINN model predictions are drawn with the SA method.
    }
    }
    \label{fig:slip_sa_bspline}
\end{figure}

\begin{figure}[t]
\centering
    \includegraphics[width=1.0\columnwidth, trim={0cm 0cm 0cm 0cm}, clip]{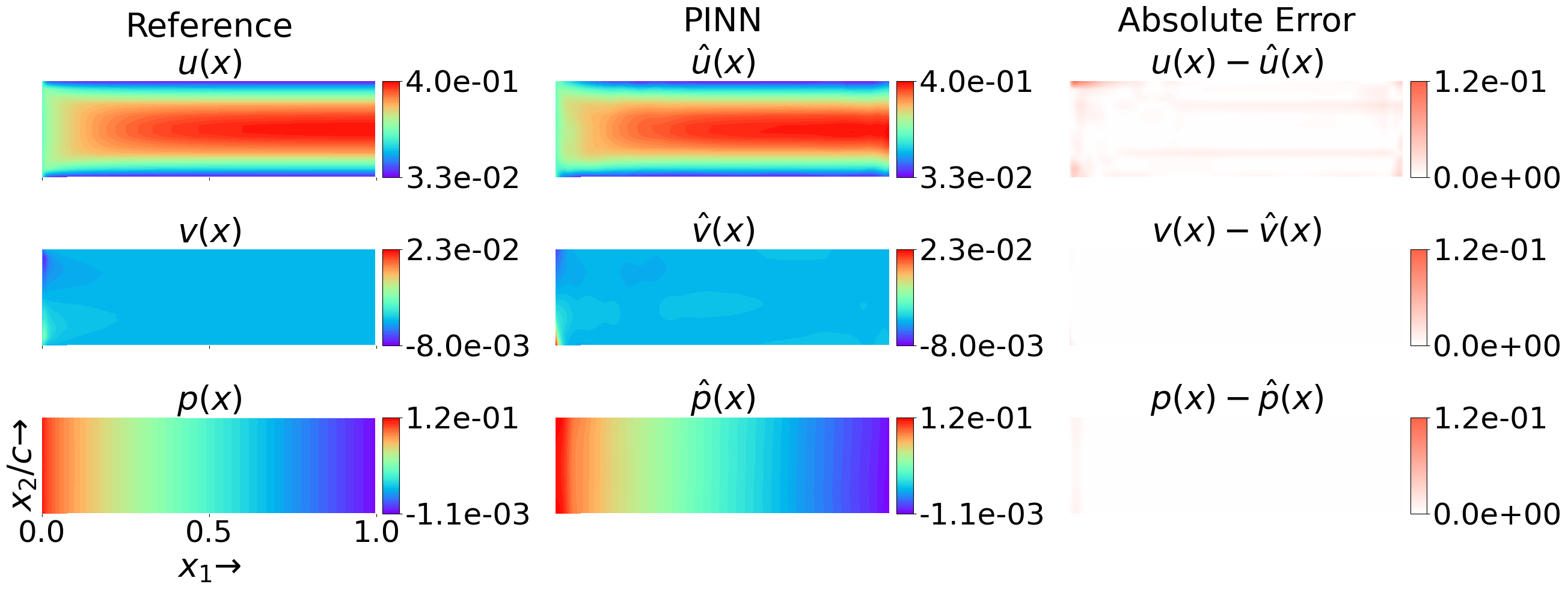} 
    \caption{ \red{Contour plots for the solution of the BFS-No-Slip problem, showing velocity components ($u_x$ and $v_y$) and pressure ($p$) values at the steady state. The PINN model predictions are drawn with the SA method. 
    }}
    \label{fig:noslip_sa_bspline}
\end{figure}

%% file: sections/6_conclusion.tex
\section{Conclusion}

In this paper, we present an improved method for PINNs that integrates SiLU activation functions and learnable B-splines to enhance the effectiveness of multi-objective loss balancing, particularly in solving fluid flow simulations governed by the Navier-Stokes equations.
To evaluate the effectiveness of our proposed solution, we conducted extensive experiments on various CFD problems, including lid-driven Cavity flow, plane Poiseuille flow, and blood flow simulations with both slip and no-slip wall boundary conditions.

We found that the choice of activation functions and the design of loss functions significantly impact the performance of loss balancing methods in diverse fluid dynamics scenarios. The effectiveness of learnable activations is dependent on the specific problem, with the most notable improvements observed in complex flows. For instance, LRA proved to be the most robust performer, achieving up to a 95.2\% improvement in complex flow scenarios when combined with learnable activations.
However, we also encountered an unexpected failure mode where GradNorm became completely unstable with B-spline+SiLU activations across all test cases, underscoring potential incompatibilities between certain loss balancing algorithms and learnable activation functions. While B-spline+SiLU networks consistently outperformed Tanh-based networks, with improvements ranging from 7.4\%-95.2\%, this enhanced accuracy is accompanied by a moderately increased computational overhead.

\red{
The computational efficiency analysis demonstrates that while B-spline+SiLU networks incur approximately 2.5 times higher parameter overhead and 4-10 times increased execution time per iteration, the accuracy improvements justify this computational cost for complex fluid flow applications. RBA emerges as the most computationally efficient adaptive method, offering a favorable trade-off between performance gains and computational overhead, while LRA, despite achieving the highest accuracy improvements, requires the most computational resources due to its gradient-intensive calculations.
}

%% file: main.bbl
\begin{thebibliography}{10}
\renewcommand*{\sfdefault}{PTSansNarrow-TLF}

\bibitem{anagnostopoulos4586276residual}
S.~J. Anagnostopoulos, J.~D. Toscano, N.~Stergiopulos, and G.~E. Karniadakis.
\newblock Residual-based attention in physics-informed neural networks.
\newblock {\em Computer Methods in Applied Mechanics and Engineering}, 421:116805, 2024.

\bibitem{ballarin2017numerical}
F.~Ballarin, E.~Faggiano, A.~Manzoni, A.~Quarteroni, G.~Rozza, S.~Ippolito, C.~Antona, and R.~Scrofani.
\newblock Numerical modeling of hemodynamics scenarios of patient-specific coronary artery bypass grafts.
\newblock {\em Biomechanics and modeling in mechanobiology}, 16:1373--1399, 2017.

\bibitem{basir2023investigating}
S.~Basir.
\newblock Investigating and mitigating failure modes in physics-informed neural networks (pinns).
\newblock {\em Communications in Computational Physics}, 33(5):1240--1269, 2023.

\bibitem{basir2022critical}
S.~Basir and I.~Senocak.
\newblock Critical investigation of failure modes in physics-informed neural networks.
\newblock In {\em AIAA SCITECH 2022 Forum}, p. 2353, 2022.

\bibitem{bischof2021multi}
R.~Bischof and M.~A. Kraus.
\newblock Multi-objective loss balancing for physics-informed deep learning.
\newblock {\em Computer Methods in Applied Mechanics and Engineering}, 439:117914, 2025.

\bibitem{brunton2017discovering}
S.~Brunton.
\newblock Discovering governing equations from data by sparse identification of nonlinear dynamics.
\newblock In {\em APS March Meeting Abstracts}, vol. 2017, pp. X49--004, 2017.

\bibitem{caro2012mechanics}
C.~G. Caro.
\newblock {\em The mechanics of the circulation}.
\newblock Cambridge University Press, 2012.

\bibitem{chen2018gradnorm}
Z.~Chen, V.~Badrinarayanan, C.-Y. Lee, and A.~Rabinovich.
\newblock Gradnorm: Gradient normalization for adaptive loss balancing in deep multitask networks.
\newblock In {\em Proceedings of the 35th International Conference on Machine Learning (ICML)}, pp. 794--803, 2018.

\bibitem{dauphin2014identifying}
Y.~N. Dauphin, R.~Pascanu, C.~Gulcehre, K.~Cho, S.~Ganguli, and Y.~Bengio.
\newblock Identifying and attacking the saddle point problem in high-dimensional non-convex optimization.
\newblock {\em Advances in neural information processing systems}, 27, 2014.

\bibitem{de1972calculating}
C.~De~Boor.
\newblock On calculating with b-splines.
\newblock {\em Journal of Approximation theory}, 6(1):50--62, 1972.

\bibitem{evans2022partial}
L.~C. Evans.
\newblock {\em Partial differential equations}, vol.~19.
\newblock American Mathematical Society, 2022.

\bibitem{farea2025:learnable}
A.~Farea and M.~S. Celebi.
\newblock {Learnable activation functions in physics-informed neural networks for solving partial differential equations}.
\newblock {\em Computer Physics Communications}, 315:109753, 2025.

\bibitem{farea2025:qcpinn}
A.~Farea, S.~Khan, and M.~S. Celebi.
\newblock {QCPINN: Quantum-Classical Physics-Informed Neural Networks for Solving PDEs}, 2025.

\bibitem{farea2025:fsi-ibm}
A.~Farea, S.~Khan, R.~Daryani, E.~C. Ersan, and M.~S. Celebi.
\newblock {Learning Fluid-Structure Interaction Dynamics with Physics-Informed Neural Networks and Immersed Boundary Methods}, 2025.

\bibitem{feigley2011deriving}
C.~E. Feigley, T.~H. Do, J.~Khan, E.~Lee, N.~D. Schnaufer, and D.~C. Salzberg.
\newblock Deriving realistic source boundary conditions for a cfd simulation of concentrations in workroom air.
\newblock {\em Annals of occupational hygiene}, 55(4):410--420, 2011.

\bibitem{Heydari2020}
A.~A. Heydari, C.~A. Thompson, and A.~Mehmood.
\newblock Softadapt: Techniques for adaptive loss weighting of neural networks with multi-part loss functions.
\newblock {\em CoRR}, abs/1912.12355, 2019.

\bibitem{jagtap2020adaptive}
A.~D. Jagtap, K.~Kawaguchi, and G.~E. Karniadakis.
\newblock Adaptive activation functions accelerate convergence in deep and physics-informed neural networks.
\newblock {\em Journal of Computational Physics}, 404:109136, 2020.

\bibitem{ji2021stiff}
W.~Ji, W.~Qiu, Z.~Shi, S.~Pan, and S.~Deng.
\newblock Stiff-pinn: Physics-informed neural network for stiff chemical kinetics.
\newblock {\em The Journal of Physical Chemistry A}, 125(36):8098--8106, 2021.

\bibitem{jin2021nsfnets}
X.~Jin, S.~Cai, H.~Li, and G.~E. Karniadakis.
\newblock Nsfnets (navier-stokes flow nets): Physics-informed neural networks for the incompressible navier-stokes equations.
\newblock {\em Journal of Computational Physics}, 426:109951, 2021.

\bibitem{karniadakis2021physics}
G.~E. Karniadakis, I.~G. Kevrekidis, L.~Lu, P.~Perdikaris, S.~Wang, and L.~Yang.
\newblock Physics-informed machine learning.
\newblock {\em Nature Reviews Physics}, 3(6):422--440, 2021.

\bibitem{kolmogorov1957representation}
A.~N. Kolmogorov.
\newblock On the representation of continuous functions of many variables by superposition of continuous functions of one variable and addition.
\newblock {\em Doklady Akademii Nauk SSSR}, 114(5):953--956, 1957.

\bibitem{krishnapriyan2021characterizing}
A.~Krishnapriyan, A.~Gholami, S.~Zhe, R.~Kirby, and M.~W. Mahoney.
\newblock Characterizing possible failure modes in physics-informed neural networks.
\newblock {\em Advances in Neural Information Processing Systems}, 34:26548--26560, 2021.

\bibitem{li2022dynamic}
S.~Li and X.~Feng.
\newblock Dynamic weight strategy of physics-informed neural networks for the 2d navier--stokes equations.
\newblock {\em Entropy}, 24(9):1254, 2022.

\bibitem{liu2021non}
B.~Liu, Z.~Liu, T.~Zhang, and T.~Yuan.
\newblock Non-differentiable saddle points and sub-optimal local minima exist for deep relu networks.
\newblock {\em Neural Networks}, 144:75--89, 2021.

\bibitem{liu2024kan}
Z.~Liu, Y.~Wang, S.~Vaidya, F.~Ruehle, J.~Halverson, M.~Solja{\v{c}}i{\'c}, T.~Y. Hou, and M.~Tegmark.
\newblock Kan: Kolmogorov-arnold networks.
\newblock {\em International Conference on Learning Representations}, 2025.

\bibitem{lobato2017multi}
F.~S. Lobato and V.~Steffen~Jr.
\newblock {\em Multi-objective optimization problems: concepts and self-adaptive parameters with mathematical and engineering applications}.
\newblock Springer, 2017.

\bibitem{maddu2022inverse}
S.~Maddu, D.~Sturm, C.~L. M{\"u}ller, and I.~F. Sbalzarini.
\newblock Inverse dirichlet weighting enables reliable training of physics-informed neural networks.
\newblock {\em Machine Learning: Science and Technology}, 3(1):015026, 2022.

\bibitem{mao2020physics}
Z.~Mao, A.~D. Jagtap, and G.~E. Karniadakis.
\newblock Physics-informed neural networks for high-speed flows.
\newblock {\em Computer Methods in Applied Mechanics and Engineering}, 360:112789, 2020.

\bibitem{mcclenny2020self}
L.~D. McClenny and U.~M. Braga-Neto.
\newblock Self-adaptive physics-informed neural networks.
\newblock {\em Journal of Computational Physics}, 474:111722, 2023.

\bibitem{moseley2023finite}
B.~Moseley, A.~Markham, and T.~Nissen-Meyer.
\newblock Finite basis physics-informed neural networks (fbpinns): a scalable domain decomposition approach for solving differential equations.
\newblock {\em Advances in Computational Mathematics}, 49(4):62, 2023.

\bibitem{pope2001turbulent}
S.~B. Pope.
\newblock Turbulent flows.
\newblock {\em Measurement Science and Technology}, 12(11):2020--2021, 2001.

\bibitem{rahaman2019spectral}
N.~Rahaman, A.~Baratin, D.~Arpit, F.~Draxler, M.~Lin, F.~Hamprecht, Y.~Bengio, and A.~Courville.
\newblock On the spectral bias of neural networks.
\newblock In {\em International Conference on Machine Learning}, pp. 5301--5310. PMLR, 2019.

\bibitem{reinbold2021robust}
P.~A. Reinbold, L.~M. Kageorge, M.~F. Schatz, and R.~O. Grigoriev.
\newblock Robust learning from noisy, incomplete, high-dimensional experimental data via physically constrained symbolic regression.
\newblock {\em Nature communications}, 12(1):3219, 2021.

\bibitem{ren2022phycrnet}
P.~Ren, C.~Rao, Y.~Liu, J.-X. Wang, and H.~Sun.
\newblock Phycrnet: Physics-informed convolutional-recurrent network for solving spatiotemporal pdes.
\newblock {\em Computer Methods in Applied Mechanics and Engineering}, 389:114399, 2022.

\bibitem{sahli2020physics}
F.~Sahli~Costabal, Y.~Yang, P.~Perdikaris, D.~E. Hurtado, and E.~Kuhl.
\newblock Physics-informed neural networks for cardiac activation mapping.
\newblock {\em Frontiers in Physics}, 8:42, 2020.

\bibitem{son2023enhanced}
H.~Son, S.~W. Cho, and H.~J. Hwang.
\newblock Enhanced physics-informed neural networks with augmented lagrangian relaxation method (al-pinns).
\newblock {\em Neurocomputing}, 548:126424, 2023.

\bibitem{sun2020surrogate}
L.~Sun, H.~Gao, S.~Pan, and J.-X. Wang.
\newblock Surrogate modeling for fluid flows based on physics-constrained deep learning without simulation data.
\newblock {\em Computer Methods in Applied Mechanics and Engineering}, 361:112732, 2020.

\bibitem{tancik2020fourier}
M.~Tancik, P.~Srinivasan, B.~Mildenhall, S.~Fridovich-Keil, N.~Raghavan, U.~Singhal, R.~Ramamoorthi, J.~Barron, and R.~Ng.
\newblock Fourier features let networks learn high frequency functions in low dimensional domains.
\newblock {\em Advances in neural information processing systems}, 33:7537--7547, 2020.

\bibitem{temam2024navier}
R.~Temam.
\newblock {\em Navier--Stokes equations: theory and numerical analysis}, vol. 343.
\newblock American Mathematical Society, 2024.

\bibitem{van2022optimally}
R.~van~der Meer, C.~W. Oosterlee, and A.~Borovykh.
\newblock Optimally weighted loss functions for solving pdes with neural networks.
\newblock {\em Journal of Computational and Applied Mathematics}, 405:113887, 2022.

\bibitem{vemuri2023gradient}
S.~K. Vemuri and J.~Denzler.
\newblock Gradient statistics-based multi-objective optimization in physics-informed neural networks.
\newblock {\em Sensors}, 23(21):8665, 2023.

\bibitem{wang2018propagation}
J.-X. Wang, C.~J. Roy, and H.~Xiao.
\newblock Propagation of input uncertainty in the presence of model-form uncertainty: a multifidelity approach for computational fluid dynamics applications.
\newblock {\em ASCE-ASME J Risk and Uncert in Engrg Sys Part B Mech Engrg}, 4(1), 2018.

\bibitem{wang2022respecting}
S.~Wang, S.~Sankaran, and P.~Perdikaris.
\newblock Respecting causality is all you need for training physics-informed neural networks.
\newblock {\em Computer Methods in Applied Mechanics and Engineering}, 2024.

\bibitem{wang2021understanding}
S.~Wang, Y.~Teng, and P.~Perdikaris.
\newblock {Understanding and Mitigating Gradient Flow Pathologies in Physics-Informed Neural Networks}.
\newblock {\em SIAM Journal on Scientific Computing}, 43(5):A3055--A3081, 2021.

\bibitem{wang2022and}
S.~Wang, X.~Yu, and P.~Perdikaris.
\newblock When and why pinns fail to train: A neural tangent kernel perspective.
\newblock {\em Journal of Computational Physics}, 449:110768, 2022.

\bibitem{xiang2022self}
Z.~Xiang, W.~Peng, X.~Liu, and W.~Yao.
\newblock Self-adaptive loss balanced physics-informed neural networks.
\newblock {\em Neurocomputing}, 496:11--34, 2022.

\bibitem{xu2018assessment}
P.~Xu, X.~Liu, H.~Zhang, D.~Ghista, D.~Zhang, C.~Shi, and W.~Huang.
\newblock Assessment of boundary conditions for cfd simulation in human carotid artery.
\newblock {\em Biomechanics and Modeling in Mechanobiology}, 17(6):1581--1597, 2018.

\bibitem{wight2020solving}
C.~L.~W. Zhao et~al.
\newblock Solving allen-cahn and cahn-hilliard equations using the adaptive physics informed neural networks.
\newblock {\em Communications in Computational Physics}, 29(3):930--954, 2021.

\end{thebibliography}
